\documentclass[11pt]{article}
\usepackage{amssymb,amsmath,amsfonts}
\usepackage{graphicx}
\usepackage{graphics}
\usepackage{eepic,epsfig}

\@addtoreset{equation}{section}

\makeatother

\begin{document}

\title{Topological effects in fermion condensate induced by cosmic string
and compactification on AdS bulk}
\author{S. Bellucci$^{1}$\thanks{%
E-mail: bellucci@lnf.infn.it },\, W. Oliveira dos Santos$^{1,2}$\thanks{%
E-mail: wagner.physics@gmail.com},\, E. R. Bezerra de Mello$^{2}$\thanks{%
E-mail: emello@fisica.ufpb.br},\, A. A. Saharian$^{3}$\thanks{%
E-mail: saharian@ysu.am} \\
\\
\textit{$^1$ INFN, Laboratori Nazionali di Frascati,}\\
\textit{Via Enrico Fermi 54, 00044 Frascati, Italy} \vspace{0.3cm}\\
\textit{$^{2}$Departamento de F\'{\i}sica, Universidade Federal da Para\'{\i}%
ba}\\
\textit{58.059-970, Caixa Postal 5.008, Jo\~{a}o Pessoa, PB, Brazil}\vspace{%
0.3cm}\\
\textit{$^3$Department of Physics, Yerevan State University,}\\
\textit{1 Alex Manoogian Street, 0025 Yerevan, Armenia}}
\maketitle

\begin{abstract}
We investigate the fermion condensate (FC) for a massive spinor field on
background of the 5-dimensional locally anti-de Sitter (AdS) spacetime with
a compact dimension and in the presence of a cosmic string carrying a
magnetic flux. The FC is decomposed into two contributions. The first one
corresponds to the geometry without compactification and the second one is
induced by the compactification. Depending on the values of the parameters,
the total FC can be either positive or negative. As a limiting case, the
expression for the FC on locally Minkowski spacetime is derived. It vanishes
for a massless fermion field and the nonzero FC on the AdS bulk in the
massless case is an effect induced by gravitation. This shows that the
gravitational field may essentially influence the parameters space for phase
transitions. For a massive field the FC diverges on the string as the
inverse cube of the proper distance from the string. In the case of a
massless field, depending on the magnetic flux along the string and planar
angle deficit, the limiting value of the FC on the string can be either
finite or infinite. At large distances, the decay of the FC as a function of
the distance from the string is power law for both cases of massive and
massless fields. For a cosmic string on the Minkowski bulk and for a massive
field the decay is exponential. The topological part in the FC vanishes on
the AdS boundary. We show that the FCs coincide for the fields realizing two
inequivalent irreducible representations of the Clifford algebra. In the
special case of the zero planar angle deficit, the results presented in this
paper describe Aharonov-Bohm-type effects induced by magnetic fluxes in
curved spacetime.
\end{abstract}

Keywords: cosmic string, anti-de Sitter spacetime, fermion condensate,
compact dimension

\bigskip

\section{Introduction}

\label{Int}

The phenomenon of Fermi condensation is important in both condensed matter
physics and quantum field theory. It plays a prominent role in the studies
of superconductivity and phase transitions, in quantum chromodynamics, in
models of dynamical mass generation and symmetry breaking. The
characteristic feature of the Fermi condensation is the appearance of
nonzero fermion condensate (FC) defined as the expectation value $\langle 
\bar{\psi}\psi \rangle $. Various mechanisms for the formation of the FC
have been considered in the literature. They include different kinds of
interactions of fermion fields, in particular, the Nambu-Jona-Lasinio-type
models with self-interacting fields. In some models the FC is related to the
gauge field condensate (gluon condensate in quantum chromodynamics). An
interesting direction in the investigations of the Fermi condensation is the
dependence of the FC on the local geometry and topology of the background
spacetime (see, for instance, \cite{Mira15}-\cite{Chu20b} and references
therein). In particular, the boundary conditions in the presence of
boundaries or periodicity conditions along compact dimensions imposed on the
fields may either reduce or enlarge the parameters space for phase
transitions. In some cases they may exclude the possibility for the
dynamical symmetry breaking.

In the present paper we study the combined effects of the gravitational
field and spatial topology on the vacuum condensate for a massive fermionic
field. As the background geometry we take a locally anti-de Sitter (AdS)
spacetime and two sources for the nontrivial topology are considered: the
presence of a cosmic string and a compactification of one of the spatial
dimensions. Our choice of the AdS spacetime is mainly motivated by two
reasons. The first one is related to its high symmetry (for geometrical
properties of the AdS spacetime and coordinate systems see \cite{Grif09})
that allows to obtain exact expressions for the physical characteristics of
the quantum vacuum. The investigation for the influence of the gravitational
field on the properties of the vacuum in these types of exactly solvable
problems will help to get an idea about the effects in more complicated
geometries, including the cosmological and black hole backgrounds. The
second motivation is related to the important role of the AdS spacetime in
recent developments of the theoretical physics. The latter include
braneworld models with large extra dimensions and AdS/CFT correspondence.
The braneworld scenario (see \cite{Maar10} for a review) was suggested as a
geometrical solution to the hierarchy problem between the electroweak and
Planckian energy scales. Most of the braneworld models are formulated in the
background of AdS spacetime and contain branes on which a part of the fields
are located. These models naturally appear in string theories and provide a
new interesting framework for discussions of various problems in particle
physics and cosmology. The second exciting development, the AdS/CFT
correspondence \cite{Ahar00,Ammo15}, is an example for the realization of
the holographic principle. It states the duality between the string or field
theories on the AdS bulk and conformal field theory on the AdS boundary.
This duality provides a unique possibility to investigate the strong
coupling effects in the theory by mapping it to the weak coupling sector of
the dual theory. Applications have been considered in string and field
theories and in strong coupling problems of condensed matter physics.

The presence of compact spatial dimensions is a general feature of a number
of models in high-energy physics, including string theories and
supergravity, as well as of field theoretical models in condensed matter
physics. The vacuum fluctuations of quantum fields along those dimensions
are quantized and, as a consequence, contributions in the vacuum expectation
values (VEVs) of physical observables appear that depend on the geometrical
characteristics of the compactification. The effects of compactification on
the properties of the quantum vacuum in locally AdS spacetime have been
considered previously (for a recent review see \cite{Saha21}). The vacuum
energy, the Casimir forces and the radion stabilization in braneworld models
with compact dimensions were discussed in \cite{Flac03a}-\cite{Fran08}. The
Wightman function, the VEVs of the field squared and energy-momentum tensor
and the induced cosmological constant on the branes for a scalar field with
general curvature coupling parameter were investigated in \cite%
{Saha06a,Saha06b,Saha06c}. The presence of extra compact subspace relax the
fine-tunings of the fundamental parameters in braneworlds. For charged
fields, among the interesting characteristic features of the
compactification is the possibility for the appearance of nonzero currents
along compact dimensions. These vacuum currents may serve as sources for
large scale cosmological magnetic fields. The VEVs of the current density in
a locally AdS spacetime with an arbitrary number of toroidally compactified
Poincar\'{e} spatial dimensions and in the presence of branes parallel to
the AdS boundary have been studied in \cite{Beze15cu,Bell15cu,Bell16cu} and 
\cite{Bell17cu,Bell18cu,Bell20cu} for charged scalar and fermionic fields,
respectively.

The second type of topological effects we are going to consider here are
induced by the presence of a cosmic string. This type of linear topological
defects are formed during symmetry-breaking phase transitions in the early
Universe \cite{Kibb76,Vile94} and lead to a number of interesting effects in
astrophysics and cosmology. The latter include the gravitational lensing,
the generation of gravitational waves, gamma ray bursts and cosmic rays.
Among the observable consequences of cosmic strings are small
non-Gaussianities in the cosmic microwave background. More recently, a
mechanism for the formation of fundamental cosmic superstrings has been
proposed within the framework of brane inflationary models (for reviews see 
\cite{Cope11,Cher15}). In the simplified model of a cosmic string, the local
geometry outside its core is not changed and the influence on the properties
of the vacuum is purely topological. This influence has been investigated
for scalar, fermionic and electromagnetic fields, mainly in locally
Minkowskian spacetime. The geometry of a cosmic string in the background of
AdS spacetime has been considered in \cite{Dehg02,Ball11,Padu16}. The vacuum
polarization by a cosmic string in AdS spacetime is investigated in \cite%
{Beze12} and \cite{Beze13} for scalar and fermionic fields, respectively.
The scalar and fermionic vacuum currents around a cosmic string carrying a
magnetic flux along its core in the background of AdS spacetime with
compactified spatial dimension have been considered in \cite{Oliv19,Bell20}.
The VEVs of the field squared and of the energy-momentum tensor in that
geometry were discussed in \cite{Oliv20}. The vacuum fermionic current
induced by a magnetic flux running along the cosmic string on AdS bulk in
the presence of a brane parallel to the AdS boundary was studied as well 
\cite{Bell21}.

This paper is organized as follows. In Section \ref{sec:Setup} we describe
the problem and present the complete set of fermionic modes. The Section \ref%
{sec:FCcs} is devoted to the investigation of the FC in the geometry without
compactification for a massive fermionic field realizing the irreducible
representation of the Clifford algebra. The cosmic string induced
contribution is explicitly extracted and various asymptotic limits are
considered. Applications are discussed in models invariant under the parity
transformation and charge conjugation. The effects of compactification of a
spatial dimension on the FC are discussed in Section \ref{sec:Comp}. The
main results of the paper are summarized in Section \ref{sec:Conc}.

\section{Setup and wave-functions}

\label{sec:Setup}

\begin{figure}[tbph]
\begin{center}
\epsfig{figure=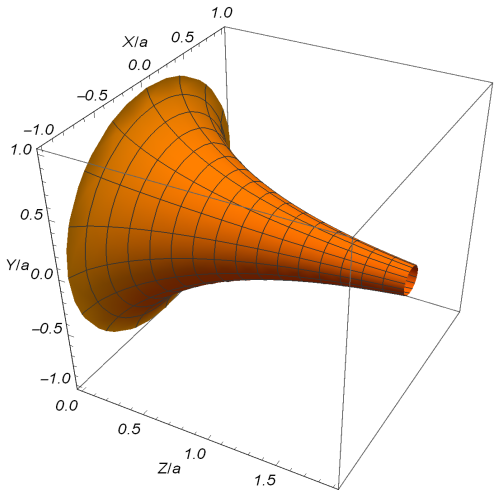,width=6.5cm,height=6.5cm} 
\end{center}
\caption{The core of the defect embedded in 3-dimensional
Euclidean space with the coordinates $(X,Y,Z)$.}
\label{fig0}
\end{figure}

We consider a simplified geometry of a cosmic string with zero thickness
core in background of a (1+4)-dimensional locally AdS spacetime described by
the line element 
\begin{equation}
ds^{2}=e^{-2y/a}\left( dt^{2}-dr^{2}-r^{2}d\phi ^{2}-dz^{2}\right) -dy^{2}\ .
\label{ds1}
\end{equation}%
For polar coordinates $(r,\phi )$ in two-dimensional subspace one has $r\geq
0$ and $\phi \in \lbrack 0,\ \phi _{0}]$. The coordinates $(t,y)$ are within
the range $-\infty <t,y<+\infty $ and we will assume that the direction
along the $z$-axis is compactified to a circle of the length $L$, so $0\leq
z\leq L$. For a fixed value of the coordinate $y$, the line element (\ref%
{ds1}) is reduced to the standard geometry of (1+3)-dimensional cosmic
string with planar angle deficit $2\pi -\phi _{0}$ compactified along its
axis. For $r>0$, the metric tensor corresponding to (\ref{ds1}) is a
solution of the Einstein equation for gravitational fields in
(1+4)-dimensional spacetime in the presence of negative cosmological
constant $\Lambda $. The latter determines the spacetime curvature scale as $%
a=\sqrt{-6/\Lambda }$. The core of the defect under consideration is given
by the 2-dimensional hypersurface $r=0$ and is covered by the coordinates $%
(z,y)$. The corresponding spatial geometry is described by the line element $%
dl_{c}^{2}=dy^{2}+e^{-2y/a}dz^{2}$. The respective constant negative
curvature 2-dimensional surface is known as Beltrami pseudosphere. In Fig. %
\ref{fig0} we have displayed the defect core embedded in a 3-dimensional
Euclidean space with the coordinates $(X,Y,Z)$. The embedding can be
realized in accordance with%
\begin{equation}
X=af(y)\cos \varphi ,\;Y=af(y)\sin \varphi ,\;Z=y-y_{0}+a\ln \left[ 1+\sqrt{%
1-f^{2}(y)}\right] -a\sqrt{1-f^{2}(y)},  \label{embed}
\end{equation}%
where $\varphi =2\pi z/L$, $f(y)=e^{-\left( y-y_{0}\right) /a}$ and $%
y_{0}=a\ln \left( L/2\pi a\right) $. Note that only the part of the core
corresponding to the region $y\geq y_{0}$ is embedded in a 3-dimensional
Euclidean space.

For $r>0$ the geometry given by (\ref{ds1}) is conformally flat. That can be
seen by introducing the Poincar\'{e} coordinate $w$ defined by $w=ae^{y/a}$.
With this coordinate, the line element is written in explicitly conformally
flat form 
\begin{equation}
ds^{2}=g_{\mu \nu }dx^{\mu }dx^{\nu }=\left( \frac{a}{w}\right) ^{2}\left(
dt^{2}-dr^{2}-r^{2}d\phi ^{2}-dw^{2}-dz^{2}\right) \ ,  \label{ds2}
\end{equation}%
where $w\in \lbrack 0,\ \infty )$. The limiting values $w=0$ and $w=\infty $
correspond to the AdS boundary and horizon, respectively. In what follows we
will work in the coordinate system $x^{\mu }=(t,r,\phi ,w,z)$ with the
metric tensor $g_{\mu \nu }=(a/w)^{2}\mathrm{diag}(1,-1,-r^{2},-1,-1)$.

We are interested in the influence of cosmic string and compactification on
the properties of the ground state for a fermionic field. In the irreducible
representation of the Clifford algebra the latter is presented by a
4-component spinor. In odd number of spacetime dimensions there are two
inequivalent irreducible representations. In (1+4)-dimensional spacetime the
corresponding set of flat spacetime Dirac matrices $\gamma _{(s)}^{(b)}$, $%
b=0,1,2,3,4$, can be constructed by adding to the set of the matrices $%
\{\gamma ^{(0)},\gamma ^{(1)},\gamma ^{(2)},\gamma ^{(3)}\}$ in
(1+3)-dimensional spacetime the matrix $s\gamma ^{(4)}=-s\gamma ^{(0)}\gamma
^{(1)}\gamma ^{(2)}\gamma ^{(3)}$. Here, $s=\pm 1$ distinguish two
inequivalent irreducible representations. Let us denote by $\psi _{(s)}$ the
4-component field that realizes the irreducible representation with a given $%
s$. The corresponding Lagrangian density reads%
\begin{equation}
L_{(s)}=\bar{\psi}_{(s)}\left[ i\gamma _{(s)}^{\mu }\left( \partial _{\mu
}+\Gamma _{\mu }^{(s)}+ieA_{\mu }\right) -m\right] \psi _{(s)},  \label{Ls}
\end{equation}%
where $\bar{\psi}_{(s)}=\psi _{(s)}^{\dagger }\gamma ^{(0)}$ is the Dirac
adjoint, $\Gamma _{\mu }^{(s)}$ is the spin connection, and $A_{\mu }$ is
the vector potential for an external gauge field. The curved spacetime Dirac
matrices are expressed in terms of the matrices $\gamma _{(s)}^{(b)}$ as $%
\gamma _{(s)}^{\mu }=e_{(b)}^{\mu }\gamma _{(s)}^{(b)}$ with vielbein fields 
$e_{(b)}^{\mu }$. Here and below, the tensorial indices $\mu =0,1,2,3,4$
correspond to the coordinates $(t,r,\phi ,w,z)$, respectively.

For the matrices $\gamma _{(s)}^{\mu }$, $\mu =0,4$, we take $\gamma
_{(s)}^{\mu }=(w/a)\delta _{b}^{\mu }\gamma _{(s)}^{(b)}$ with 
\begin{equation}
\gamma ^{(0)}=\left( {%
\begin{array}{cc}
0 & -i \\ 
i & 0%
\end{array}%
}\right) \ ,\;\gamma _{(+1)}^{(4)}=-\gamma _{(-1)}^{(4)}=\gamma
^{(4)}=\left( {%
\begin{array}{cc}
0 & i \\ 
i & 0%
\end{array}%
}\right) .  \label{Dirac04}
\end{equation}%
For the construction of the remaining matrices we use the 3D flat space $%
2\times 2$ Pauli matrices $\sigma ^{1}$, $\sigma ^{2}$, $\sigma ^{3}$ in the
cylindrical coordinate system $(r,\phi ,w)$: 
\begin{equation}
\sigma ^{l}=\left( \frac{i}{r}\right) ^{l-1}\left( {%
\begin{array}{cc}
0 & (-1)^{l-1}e^{-iq\phi } \\ 
e^{iq\phi } & 0%
\end{array}%
}\right) \ ,  \label{PauliCyl}
\end{equation}%
for $l=1,2$ and $\sigma ^{3}=\mathrm{diag}(1,-1)$. Here and in what follows
we use the notation $q=2\pi /\phi _{0}$. In terms of those matrices we take 
\cite{Bell20} 
\begin{equation}
\gamma _{(s)}^{l}=-i(w/a)\mathrm{diag}(\sigma ^{l},-\sigma ^{l}),\;l=1,2,3.
\label{gaml}
\end{equation}%
Below, the simplified notations $\gamma ^{\mu }\equiv \gamma _{(+1)}^{\mu }$
and $\Gamma _{\mu }=\Gamma _{\mu }^{(+1)}$ will be used. The product $\gamma
_{(s)}^{\mu }\Gamma _{\mu }^{(s)}$ in the Lagrangian density (\ref{Ls}) is
presented as $\gamma _{(s)}^{\mu }\Gamma _{\mu }^{(s)}=-2\gamma
^{3}/w+(1-q)\gamma ^{1}/(2r)$ and it is the same for both the
representations.

The investigation of the FC for the fields with $s=+1$ and $s=-1$ can be
unified by passing to a new representation $\psi _{(s)}^{\prime }$ with $%
\psi _{(+1)}^{\prime }=\psi _{(+1)}$ and $\psi _{(-1)}^{\prime }=-\gamma
^{(4)}\psi _{(-1)}$, where the $\gamma ^{(4)}$ matrix is given by (\ref%
{Dirac04}). By using the relations $\gamma ^{(4)}\gamma _{(-1)}^{\mu }\gamma
^{(4)}=\gamma ^{\mu }$ and $\gamma ^{(4)}\gamma _{(s)}^{\mu }\Gamma _{\mu
}^{(s)}\gamma ^{(4)}=\gamma ^{\mu }\Gamma _{\mu }$, the Lagrangian density
for the new fields is presented as%
\begin{equation}
L_{(s)}=\bar{\psi}_{(s)}^{\prime }\left[ i\gamma ^{\mu }\left( \partial
_{\mu }+\Gamma _{\mu }+ieA_{\mu }\right) -sm\right] \psi _{(s)}^{\prime },
\label{Lsp}
\end{equation}%
and differs by the sign of the mass term for two irreducible
representations. The investigation described below will be presented in the
representation (\ref{Lsp}) and the final result will be translated for the
initial fields $\psi _{(s)}$.

The equation of motion corresponding to the Lagrangian density (\ref{Lsp})
(omitting the prime and the index $(s)$) reads 
\begin{equation}
i\gamma ^{\mu }\left( \partial _{\mu }+\Gamma _{\mu }+ieA_{\mu }\right) \psi
-sm\psi =0\,.  \label{DiracEq}
\end{equation}%
The coordinate $z$ is compactified and in addition to the field equation one
needs to specify the periodicity condition on the field operator along that
direction. We will consider the quasi-periodicity condition 
\begin{equation}
\psi (t,r,\phi ,w,z+L)=e^{2\pi i\beta }\psi (t,r,\phi ,w,z),  \label{QPC}
\end{equation}%
with a constant parameter $\beta $ determining the change in the phase of
the transformed field. The special cases of untwisted and twisted fermionic
fields correspond to $\beta =0$ and $\beta =1/2$, respectively.

In this paper, we will consider a simple configuration of the gauge field
with the vector potential $A_{\mu }=(0,0,A_{2},0,A_{4})$ with constant
covariant components $A_{2}$ and $A_{4}$. The component $A_{2}$ is expressed
in terms of the magnetic flux $\Phi _{\mathrm{s}}$ running along the
string's core as $A_{2}=-q\Phi _{\mathrm{s}}/(2\pi )$. We can also introduce
a magnetic flux $\Phi _{\mathrm{c}}$ related to the component $A_{4}$ as $%
A_{4}=-\Phi _{\mathrm{c}}/L$. Formally, the latter can be interpreted as a
magnetic flux enclosed by the compact dimension. It gets a real physical
meaning in the braneworld realization of the model, where the setup under
consideration is embedded in a (1+5)-dimensional spacetime as a fermionic
field localized on a hypersurface (brane).

We are interested in the influence of the gravitational field and of
nontrivial spatial topology on the local properties of the fermionic vacuum.
As important local characteristic the FC will be considered. Another local
characteristic, the VEV of the current density, has been investigated in a
recent publication \cite{Bell20}. The VEV of a physical observable bilinear
in the field operator is expressed in terms of the mode sum over a complete
set of solutions of the field equation. This set of solutions for the
problem under consideration is given in \cite{Bell20} and we will describe
it to fix the notations and for the further use in the evaluation of the FC.

In accordance with the problem symmetry, the dependence of the positive and
negative energy spinorial modes on the coordinates $t$ and $z$ can be taken
in the form $e^{ik_{z}z\mp iEt}$, where $E>0$. Decomposing the $4$-component
spinor $\psi $ into the upper and lower $2$-component ones, from the Dirac
equation (\ref{DiracEq}) a second order differential equation is obtained
for each component. Separating the variables it can be seen the dependence
on the coordinates $r$ and $w$ is expressed in terms of cylinder functions.
Additional conditions relating two components of the spinor are imposed in
order to uniquely specify the wave-functions \cite{Bordag}. Denoting by $%
\sigma $ the complete set of quantum numbers specifying the solutions, the
positive and negative energy mode functions are presented as 
\begin{equation}
\psi _{\sigma }^{(\pm )}(x)=C_{\sigma }^{(\pm )}w^{5/2}\left( {%
\begin{array}{c}
J_{\beta _{j}}(\lambda r)J_{\nu _{1}}(pw)e^{-iq\phi /2} \\ 
\mp s\epsilon _{j}\kappa _{\eta }b_{\eta }^{(\pm )}J_{\beta _{j}+\epsilon
_{j}}(\lambda r)J_{\nu _{2}}(pw)e^{iq\phi /2} \\ 
is\kappa _{\eta }J_{\beta _{j}}(\lambda r)J_{\nu _{2}}(pw)e^{-iq\phi /2} \\ 
\pm i\epsilon _{j}b_{\eta }^{(\pm )}J_{\beta _{j}+\epsilon _{j}}(\lambda
r)J_{\nu _{1}}(pw)e^{iq\phi /2}%
\end{array}%
}\right) e^{iqj\phi +ik_{z}z\mp iEt}\ ,  \label{wfunc}
\end{equation}%
where $0\leq \lambda ,p<\ \infty $, $j=\pm 1/2,\pm 3/2,...$, and $J_{\nu }(x)
$ is the Bessel function. The eigenvalues of the momentum $k_{z}$ along the
axis $z$ are determined by the quasiperiodicity condition (\ref{QPC}) and
are given by $k_{z}=k_{l}=2\pi (l+\beta )/L$, where $l=0,\pm 1,\pm 2,...$.
The energy $E$ is expressed in terms of the quantum numbers by the relation 
\begin{equation}
E=\sqrt{\lambda ^{2}+p^{2}+\tilde{k}_{l}^{2}}\,\ ,  \label{E}
\end{equation}%
where%
\begin{equation}
\tilde{k}_{l}=2\pi (l+\tilde{\beta})/L,\;\tilde{\beta}=\beta -\Phi _{\mathrm{%
c}}/\Phi _{0},  \label{klt}
\end{equation}%
being $\Phi _{0}=2\pi /e$ the magnetic flux quantum. The notations appearing
in the orders of the Bessel functions are given by 
\begin{equation}
\beta _{j}=q|j+\alpha |-\epsilon _{j}/2\ ,\ \ \nu _{l}=ma+(-1)^{l}s/2\ ,
\label{order_Bessel}
\end{equation}%
where $\alpha =eA_{2}/q=-\Phi _{\mathrm{s}}/\Phi _{0}$ and $\epsilon _{j}=1$
for $j>-\alpha $ and $\epsilon _{j}=-1$ for $j<-\alpha $. The coefficients $%
\kappa _{\eta }$ and $b_{\eta }^{(\pm )}$, with $\eta =-1,+1$, are defined
by the relations 
\begin{eqnarray}
&&\kappa _{\eta }=\frac{1}{p}\left( -\tilde{k}_{l}+\eta \sqrt{\tilde{k}%
_{l}^{2}+p^{2}}\right) ,  \notag \\
&&b_{\eta }^{(\pm )}=\frac{1}{\lambda }\left( E\mp \eta \sqrt{\tilde{k}%
_{l}^{2}+p^{2}}\right) \ .  \label{kb}
\end{eqnarray}%
The normalization coefficient is given by the expression 
\begin{equation}
|C_{\sigma }^{(\pm )}|^{2}=\frac{\eta qp^{2}\lambda ^{2}}{8\pi a^{4}LE\kappa
_{\eta }b_{\eta }^{(\pm )}\sqrt{\tilde{k}_{l}^{2}+p^{2}}}\ .  \label{coeff}
\end{equation}%
The complete set of quantum numbers $\sigma $ is specified as $\sigma
=(\lambda ,p,j,l,\eta )$. Note that the mode functions (\ref{wfunc}) are the
eigenfunctions of the projection of the total angular momentum on the $w$%
-axis with the eigenvalues $qj$: 
\begin{equation}
\hat{J}_{3}\psi _{\sigma }^{(\pm )}(x)=\left( -i\partial _{\phi }+q\Sigma
^{3}/2\right) \psi _{\sigma }^{(\pm )}(x)=qj\psi _{\sigma }^{(\pm )}(x)\ ,
\label{J}
\end{equation}%
where $\Sigma ^{3}=\mathrm{diag}(\sigma ^{3},\sigma ^{3})$.

In the discussion above we have used the coordinates $(t,r,\phi ,w,z)$. The
corresponding coordinates in pure AdS spacetime with $-\infty <z<+\infty $
are referred to as Poincar\'{e} coordinates. Our choice of those coordinates
is motivated by the fact that the braneworld models and the discussions of
AdS/CFT correspondence employ the Poincar\'{e} patch. With $0<w<+\infty $,
the Poincar\'{e} coordinates cover a half of global AdS spacetime. The
second half is covered by the coordinates $(t,r,\phi ,w,z)$ with $-\infty
<w<0$. The mode functions corresponding to that patch are obtained from (\ref%
{wfunc}) by an analytic continuation. The latter is reduced to the analytic
continuation of the Bessel functions with the arguments $pw$. In a similar
way, the expressions for the FC in the second Poincar\'{e} patch are
obtained by \ a simple analytic continuation from the region $w>0$ to the
region $w<0$. In both Poincar\'{e} and global coordinates time-like Killing
vectors are present and we can define the corresponding vacuum states. It is
important to note that the Poincar\'{e} and global vacuum states are
equivalent (see, for example, \cite{Dani99,Spra99}).

\section{Fermionic condensate in the uncompactified geometry}

\label{sec:FCcs}

The FC is defined as the VEV $\langle 0|\bar{\psi}\psi |0\rangle \equiv
\langle \bar{\psi}\psi \rangle $, where $|0\rangle $ is the vacuum state
(Poincar\'{e} vacuum) and $\bar{\psi}=\psi ^{\dagger }\gamma ^{(0)}$ is the
Dirac adjoint. Note that in the definition of the Dirac adjoint $\gamma
^{(0)}$ is the flat spacetime matrix (\ref{Dirac04}). We start our
investigation first considering the geometry where the $z$-direction is not
compactified, $-\infty <z<+\infty $. The fermionic vacuum polarization in
the geometry of a straight cosmic string on the Minkowski bulk has been
investigated in \cite{Frol87}-\cite{Beze06}. The effects induced by the
presence of additional boundaries were discussed in \cite{Beze08f,Beze13f}.
The FC and the expectation values of the current density and energy-momentum
tensor in (1+2)-dimensional conical spacetime with circular boundaries have
been considered in \cite{Beze10f3}-\cite{Saha21f3}.

In the geometry with uncompactified $z$-direction the corresponding
fermionic mode functions are given by (\ref{wfunc}) where now the momentum
along the $z$-direction is continuous, $-\infty <k_{z}<+\infty $, and in the
relations (\ref{E})-(\ref{coeff}) the replacement $\tilde{k}_{l}\rightarrow
k_{z}$ should be made. In addition, in the normalization coefficient (\ref%
{coeff}) one needs to replace $L$ by $2\pi $. Expanding the field operator
in terms of the complete set $\{\psi _{\sigma }^{(-)},\psi _{\sigma
}^{(+)}\} $, and using the anticommutation relations for the creation and
annihilation operators, for the FC the following mode sum formula is
obtained: 
\begin{equation}
\langle \bar{\psi}\psi \rangle =-\frac{1}{2}\sum_{\sigma }\sum_{\chi
=-,+}\chi \bar{\psi}_{\sigma }^{(\chi )}\psi _{\sigma }^{(\chi )},
\label{FC}
\end{equation}%
where the summation goes over the set of quantum numbers as 
\begin{equation}
\sum_{\sigma }=\sum_{j}\int_{0}^{\infty }d\lambda \int_{0}^{\infty
}dp\int_{-\infty }^{\infty }dk_{z}\sum_{\eta =\pm 1}\ .  \label{Sumsig}
\end{equation}%
Here and in what follows $\sum_{j}=\sum_{j=\pm 1/2,\pm 3/2,\cdots }$. The
operators in the definition of the FC are given at the same spacetime point
and the expression in the right-hand side of (\ref{FC}) is divergent.
Various regularization schemes can be employed to make the expression
finite. For example, we can use the point-splitting technique or a cutoff
function can be introduced. The details of the evaluation procedure
described below do not depend on the specific regularization method.

Substituting the mode functions in (\ref{FC}), for the FC in the
uncompactified geometry we obtain%
\begin{equation}
\langle \bar{\psi}\psi \rangle _{\mathrm{cs}}^{\mathrm{AdS}}=-\frac{sqw^{5}}{%
16\pi ^{2}a^{4}}\sum_{\sigma }\sum_{\chi =-,+}\frac{\chi \eta p^{2}\lambda
^{2}}{E\sqrt{p^{2}+k_{z}^{2}}}J_{\nu _{1}}(pw)J_{\nu _{2}}(pw)\left[ b_{\eta
}^{(-\chi )}J_{\beta _{j}}^{2}(\lambda r)-b_{\eta }^{(\chi )}J_{\beta
_{j}+\epsilon _{j}}^{2}(\lambda r)\right] ,  \label{FC2}
\end{equation}%
where the property $b_{\eta }^{(+)}b_{\eta }^{(-)}=1$ has been used. By
taking into account the expression for $b_{\eta }^{(\pm )}$ and summing over 
$\eta $, this formula is simplified to 
\begin{eqnarray}
\langle \bar{\psi}\psi \rangle _{\mathrm{cs}}^{\mathrm{AdS}} &=&-\frac{%
sqw^{5}}{4\pi ^{2}a^{4}}\sum_{j}\int_{0}^{\infty }d\lambda \int_{0}^{\infty
}dp\int_{-\infty }^{\infty }dk_{z}\frac{\lambda p^{2}}{\sqrt{\lambda
^{2}+p^{2}+k_{z}^{2}}}  \notag \\
&&\times J_{\nu _{1}}(pw)J_{\nu _{2}}(pw)\left[ J_{\beta _{j}}^{2}(\lambda
r)+J_{\beta _{j}+\epsilon _{j}}^{2}(\lambda r)\right] \ .  \label{AdScs}
\end{eqnarray}%
From here it follows that the FC has opposite signs for the fields with $%
s=+1 $ and $s=-1$. So, in this section we will consider the case $s=+1$ and,
hence, in the corresponding expressions $\nu _{1}=ma-1/2$ and $\nu
_{2}=ma+1/2$.

For the further transformation of the expression (\ref{AdScs}), we use the
identity 
\begin{equation}
\frac{1}{\sqrt{\lambda ^{2}+p^{2}+k_{z}^{2}}}=\frac{2}{\sqrt{\pi }}%
\int_{0}^{\infty }d\tau e^{-\tau ^{2}(\lambda ^{2}+p^{2}+k_{z}^{2})}\ .
\label{first-id}
\end{equation}%
Plugging this in (\ref{AdScs}) the integrals over the quantum numbers $%
\lambda $, $p$ and $k_{z}$ are evaluated by using the results from \cite%
{Grad}. After some intermediate steps, introducing instead of $\tau $ a new
integration variable $y=r^{2}/2\tau ^{2}$, we get 
\begin{eqnarray}
\langle \bar{\psi}\psi \rangle _{\mathrm{cs}}^{\mathrm{AdS}} &=&-\frac{%
q(w/r)^{6}}{4\pi ^{2}a^{4}}\int_{0}^{\infty }dx\,x^{2}e^{-(1+\rho ^{-2})x}%
\left[ I_{\nu _{1}}\left( x/\rho ^{2}\right) -I_{\nu _{2}}\left( x/\rho
^{2}\right) \right]  \notag \\
&&\times \sum_{j}\left[ I_{\beta _{j}}\left( x\right) +I_{\beta
_{j}+\epsilon _{j}}\left( x\right) \right] \ ,  \label{FCcs1}
\end{eqnarray}%
where $I_{\nu }(z)$ represents the modified Bessel function \cite{Abra}. In (%
\ref{FCcs1}), 
\begin{equation}
\rho =r/w  \label{rho}
\end{equation}%
is the proper distance from the string, $r_{p}=ar/w$, measured in units of $%
a $.

At this point it is convenient to decompose the parameter $\alpha $ as 
\begin{equation}
\alpha =\alpha _{0}+n_{0}\ ,\quad |\alpha _{0}|<1/2\ ,  \label{alfdec}
\end{equation}%
where $n_{0}$ is an integer. Note that if we shift $j+n_{0}\rightarrow j$,
the VEV (\ref{FCcs1}) remains unchanged, which implies that it does not
depend on the integer part $n_{0}$. An integral representation for the part 
\begin{equation}
{\mathcal{J}}(q,\alpha _{0},x)=\sum_{j}\left[ I_{\beta _{j}}\left( x\right)
+I_{\beta _{j}+\epsilon _{j}}\left( x\right) \right]  \label{Jcal}
\end{equation}%
is found by using the representation for the series $\sum_{j}I_{\beta
_{j}}(x) $ given in \cite{Beze10f3}. The representation reads%
\begin{eqnarray}
{\mathcal{J}}(q,\alpha _{0},x) &=&\frac{2}{q}e^{x}+\frac{4}{\pi }%
\int_{0}^{\infty }du\frac{h(q,\alpha _{0},2u)\sinh u}{\cosh (2qu)-\cos (q\pi
)}e^{-x\cosh 2u}  \notag \\
&&+\frac{4}{q}\sideset{}{'}{\sum}_{k=1}^{[q/2]}(-1)^{k}\cos (\pi k/q)\cos
(2\pi k\alpha _{0})e^{x\cos (2\pi k/q)}\ ,  \label{J-function}
\end{eqnarray}%
where $[q/2]$ stands for the integer part of $q/2$ and 
\begin{eqnarray}
h(q,\alpha _{0},x) &=&\cos [\pi q(1/2+\alpha _{0})]\sinh [(1/2-\alpha
_{0})qx]  \notag \\
&&+\cos [\pi q(1/2-\alpha _{0})]\sinh [(1/2+\alpha _{0})qx]\ .
\label{h-function}
\end{eqnarray}%
The prime on the summation sign over $k$ means that for even values of $q$
the term with $k=q/2$ should be halved. In the case $1\leqslant q<2$, the
last term on the right-hand side of (\ref{J-function}) must be omitted. Note
that $h(q,\alpha _{0},z)$ and ${\mathcal{J}}(q,\alpha _{0},y)$ are even
functions of $\alpha _{0}$.

Introducing a new integration variable, the contribution to the condensate (%
\ref{FCcs1}) coming from the term with $e^{x}$ in (\ref{J-function}) is
presented as 
\begin{equation}
\langle \bar{\psi}\psi \rangle ^{\mathrm{AdS}}=-\frac{1}{2\pi ^{2}a^{4}}%
\int_{0}^{\infty }dx\,x^{2}e^{-x}\left[ I_{\nu _{1}}\left( x\right) -I_{\nu
_{2}}\left( x\right) \right] .  \label{FCAdS}
\end{equation}%
It does not depend on $q$ and $\alpha _{0}$ and corresponds to th FC in
(1+4)-dimensional AdS spacetime in the absence of cosmic string. As we could
expect from the maximal symmetry of the AdS spacetime the latter does not
depend on spacetime point. It is of interest to compare the FC (\ref{FCAdS})
with the condensate in de Sitter (dS) spacetime. The latter has been
investigated in \cite{Saha21dS} for the Bunch-Davies vacuum state in general
number of spatial dimensions. Specified to the case of (1+4)-dimensional dS
spacetime with the curvature radius $a$, the unregularized FC is expressed as 
\begin{equation}
\langle \bar{\psi}\psi \rangle ^{\mathrm{dS}}=\frac{1}{\pi ^{3}a^{4}}%
\int_{0}^{\infty }dx\,x^{2}e^{x}\mathrm{Im}\left[ K_{1/2-ima}\left( x\right) %
\right] ,  \label{FCdS}
\end{equation}%
where $K_{\nu }(x)$ is the Macdonald function. The expressions in the
right-hand sides of (\ref{FCAdS}) and (\ref{FCdS}) are divergent and need a
regularization with further renormalization. For the regularization of (\ref%
{FCdS}) in \cite{Saha21dS} a cutoff function is introduced. The
renormalization ambiguity is fixed by an additional condition, requiring $%
\lim_{m\rightarrow \infty }\langle \bar{\psi}\psi \rangle ^{\mathrm{dS}}=0$.
The renormalized condensate $\langle \bar{\psi}\psi \rangle _{\mathrm{ren}}^{%
\mathrm{dS}}$ is negative in spatial dimensions 3,5,6 and positive in
4-dimensional space. The renormalization of the condensate (\ref{FCAdS}) is
done in a way similar to that used in \cite{Saha21dS}. We will address that
point elsewhere.

In the present paper we are interested in the effects induced by cosmic
string. The corresponding contribution to the FC is given by 
\begin{equation}
\langle \bar{\psi}\psi \rangle _{\mathrm{cs}}=\langle \bar{\psi}\psi \rangle
_{\mathrm{cs}}^{\mathrm{AdS}}-\langle \bar{\psi}\psi \rangle ^{\mathrm{AdS}}.
\label{FC_CS_Ren}
\end{equation}%
An important point to mention here is that for $r>0$ the difference (\ref%
{FC_CS_Ren}) is finite and the regularization implicitly assumed before can
be safely removed. The physical reason of the absence of divergences in $%
\langle \bar{\psi}\psi \rangle _{\mathrm{cs}}$ is that the local geometry in
the region $r>0$ is not changed by the cosmic string and, hence, new
divergences will not arise. Substituting (\ref{J-function}) in (\ref{FCcs1})
we get%
\begin{eqnarray}
\langle \bar{\psi}\psi \rangle _{\mathrm{cs}} &=&-\frac{1}{\pi ^{2}a^{4}}%
\int_{0}^{\infty }dx\,x^{2}e^{-x}\left[ I_{\nu _{1}}\left( x\right) -I_{\nu
_{2}}\left( x\right) \right]  \notag \\
&&\times \Biggl\{\sideset{}{'}{\sum}_{k=1}^{[q/2]}(-1)^{k}\cos (\pi k/q)\cos
(2\pi k\alpha _{0})e^{-2x\rho ^{2}\sin ^{2}(\pi k/q)}  \notag \\
&&+\frac{q}{\pi }\int_{0}^{\infty }du\frac{h(q,\alpha _{0},2u)\sinh u}{\cosh
(2qu)-\cos (q\pi )}e^{-2x\rho ^{2}\cosh ^{2}u}\Biggr\}\ .  \label{FCcs0}
\end{eqnarray}%
The integral over $x$ is valuated by using the integration formula from \cite%
{Grad}: 
\begin{equation}
\int_{0}^{\infty }dx\,x^{\mu -1}e^{-ux}I_{\nu }(x)=\sqrt{\frac{2}{\pi }}%
\frac{e^{-(\mu -1/2)\pi i}Q_{\nu -1/2}^{\mu -1/2}(u)}{(u^{2}-1)^{\mu /2-1/4}}%
\ ,  \label{function_Q}
\end{equation}%
where $u>1$ and $Q_{\nu }^{\mu }(z)$ represents the associated Legendre
function \cite{Abra}. After some intermediate steps, the expression for the
FC induced by the cosmic string reads 
\begin{eqnarray}
\langle \bar{\psi}\psi \rangle _{\mathrm{cs}} &=&-\frac{\sqrt{2}}{\pi
^{5/2}a^{4}}\Biggl[\sideset{}{'}{\sum}_{k=1}^{[q/2]}(-1)^{k}\cos (\pi
k/q)\cos (2\pi k\alpha _{0}){\mathcal{Z}}_{ma}(u_{k})  \notag \\
&&+\frac{q}{\pi }\int_{0}^{\infty }dx\frac{h(q,\alpha _{0},2x)\sinh x}{\cosh
(2qx)-\cos (q\pi )}{\mathcal{Z}}_{ma}(u_{x})\Biggr]\ .  \label{FC-cs}
\end{eqnarray}%
Here we have introduced the notation 
\begin{equation}
{\mathcal{Z}}_{ma}(u)=F_{\nu _{1}}(u)-F_{\nu _{2}}(u)\,,  \label{Z_function}
\end{equation}%
with the function%
\begin{equation}
F_{\nu }(u)=\frac{e^{-i5\pi /2}Q_{\nu -1/2}^{5/2}(u)}{(u^{2}-1)^{5/4}}\,,
\label{Fnu}
\end{equation}%
and the variables 
\begin{eqnarray}
u_{k} &=&1+2\rho ^{2}\sin ^{2}(\pi k/q)\ ,  \notag \\
u_{x} &=&1+2\rho ^{2}\cosh ^{2}x\ .  \label{args-cs}
\end{eqnarray}%
Note that the FC $\langle \bar{\psi}\psi \rangle _{\mathrm{cs}}$ depends on
the coordinates $r$ and $w$ through the ratio (\ref{rho}). This property is
a consequence of the maximal symmetry of the AdS spacetime.

For a massless field, the function ${\mathcal{Z}}_{ma}(u)$ is expressed in
terms of elementary functions: 
\begin{equation}
{\mathcal{Z}}_{0}(u)=\frac{3\sqrt{\pi }}{4(1+u)^{5/2}}\ .  \label{massless-Z}
\end{equation}%
Taking this expression into (\ref{FC-cs}), we get 
\begin{eqnarray}
\langle \bar{\psi}\psi \rangle _{\mathrm{cs}} &=&-\frac{3}{16\pi ^{2}a^{4}}%
\Biggl[\sideset{}{'}{\sum}_{k=1}^{[q/2]}(-1)^{k}\frac{\cos (\pi k/q)\cos
(2\pi k\alpha _{0})}{(1+\rho ^{2}\sin ^{2}(\pi k/q))^{5/2}}  \notag \\
&&+\frac{q}{\pi }\int_{0}^{\infty }dx\frac{\sinh x}{\cosh (2qx)-\cos (q\pi )}%
\frac{h(q,\alpha _{0},2x)}{(1+\rho ^{2}\cosh ^{2}x)^{5/2}}\Biggr]\ .
\label{FCm0}
\end{eqnarray}%
Near the string, $\rho \ll 1$, and for 
\begin{equation}
2|\alpha _{0}|<1-1/q\,,  \label{cond1}
\end{equation}%
the leading term in the expansion of the FC (\ref{FCm0}) is obtained
directly putting $\rho =0$. In this case the FC is finite on the string and%
\begin{equation}
\langle \bar{\psi}\psi \rangle _{\mathrm{cs}}|_{r=0}=-\frac{3h_{0}(q,\alpha
_{0})}{16\pi ^{2}a^{4}}.  \label{FCm0r0}
\end{equation}%
Here and in what follows the notation%
\begin{equation}
h_{n}(q,\alpha _{0})=\sideset{}{'}{\sum}_{k=1}^{[q/2]}\frac{(-1)^{k}\cos
(\pi k/q)}{\sin ^{n}(\pi k/q)}\cos (2\pi k\alpha _{0})+\frac{q}{\pi }%
\int_{0}^{\infty }dx\,\frac{h(q,\alpha _{0},2x)\sinh x}{\left[ \cosh
(2qx)-\cos (q\pi )\right] \cosh ^{n}x}\ ,  \label{h_mu}
\end{equation}%
is introduced. For 
\begin{equation}
2|\alpha _{0}|>1-1/q\,,  \label{cond2}
\end{equation}%
the dominant contribution to the FC (\ref{FCm0}) comes from the integral
term. In this case we cannot directly put $\rho =0$ because the integral
diverges at the lower limit. It can be seen that the dominant contribution
to the integral comes from the integration range $x\sim \ln (1/\rho )$. On
the base of this we can show that, to the leading order,%
\begin{equation}
\langle \bar{\psi}\psi \rangle _{\mathrm{cs}}\approx -\frac{q\cos [\pi
q(1/2-|\alpha _{0}|)]}{16\pi ^{7/2}a^{4}\left( \rho /2\right) ^{1+\left(
2|\alpha _{0}|-1\right) q}}\Gamma \left( 2+\left( 1/2-|\alpha _{0}|\right)
q\right) \Gamma \left( 1/2-\left( 1/2-|\alpha _{0}|\right) q\right) ,
\label{FCcsNear}
\end{equation}%
and the condensate (\ref{FCm0}) diverges on the string like $1/\rho
^{1+\left( 2|\alpha _{0}|-1\right) q}$. Note that under the condition (\ref%
{cond2}) the leading term (\ref{FCcsNear}) is negative.

Another special case corresponds to a magnetic flux in the absence of planar
angle deficit with $q=1$. Denoting by $\langle \bar{\psi}\psi \rangle _{%
\mathrm{mf}}=\langle \bar{\psi}\psi \rangle _{\mathrm{cs}}|_{q=1}$ the part
in the FC induced by the magnetic flux, from (\ref{FC-cs}) we obtain%
\begin{equation}
\langle \bar{\psi}\psi \rangle _{\mathrm{mf}}=-\frac{\sqrt{2}\sin (\pi
\alpha _{0})}{\pi ^{7/2}a^{4}}\int_{0}^{\infty }dx\sinh (2\alpha _{0}x)\tanh
(x){\mathcal{Z}}_{ma}(1+2\rho ^{2}\cosh ^{2}x).\   \label{FCmf}
\end{equation}%
The expression for the function (\ref{h_mu}), specified for this case, takes
the form 
\begin{equation}
h_{n}(1,\alpha _{0})=\frac{1}{\pi }\sin (\pi \alpha _{0})\int_{0}^{\infty
}dx\,\frac{\sinh (2\alpha _{0}x)}{\cosh ^{n+1}x}\sinh x\ .  \label{hn1}
\end{equation}%
This expression should be used in the asymptotic estimates below for the
magnetic flux-induced contribution in the FC when the planar angle deficit
is absent.

Now, let us return to the general case of the parameter $q$ and analyse some
asymptotic properties of the FC induced by the cosmic string. In the
Minkowskian limit, we have $a\rightarrow \infty $ with $y$ fixed. This
implies in $w\approx a+y$ and $ma\gg 1$. In this limit one needs the
asymptotic behavior of the function $Q_{\nu -1/2}^{5/2}(u)$ for $u-1\ll 1$
and $\nu \gg 1$. In the literature we could find only the leading term in
the corresponding asymptotic expansion. The leading term is cancelled in the
corresponding expansion of the function ${\mathcal{Z}}_{ma}(u)$ and the
next-to-leading order term is required. In our calculation we have used the
representation (\ref{FCcs1}) with the function ${\mathcal{J}}(q,\alpha
_{0},y)$ from (\ref{J-function}). By using the uniform asymptotic expansion
for the function $I_{\nu }(x)$ for large values of the order it can be seen
that in the limit $a\rightarrow \infty $, to the leading order, one finds 
\begin{equation}
\int_{0}^{\infty }dxx^{2}e^{-\left( 1+2b^{2}/w^{2}\right) x}\left[
I_{ma-1/2}(x)-I_{ma+1/2}(x)\right] \approx \sqrt{\frac{2}{\pi }}%
a^{4}m^{4}f_{3/2}\left( 2mb\right) ,  \label{AsMink}
\end{equation}%
with the notation $f_{\nu }(x)=K_{\nu }(x)/x^{\nu }$, being $K_{\nu }(x)$
the Macdonald function \cite{Abra}. For the FC induced by cosmic string in
(1+4)-dimensional Minkowski spacetime this gives 
\begin{eqnarray}
\langle \bar{\psi}\psi \rangle _{\mathrm{cs}}^{\mathrm{(M)}} &=&-\frac{\sqrt{%
2}m^{4}}{\pi ^{5/2}}\left[ \frac{q}{\pi }\int_{0}^{\infty }dx\frac{%
h(q,\alpha _{0},2x)\sinh x}{\cosh (2qx)-\cos (q\pi )}f_{3/2}\left( 2mr\cosh
x\right) \right.  \notag \\
&&\left. +\sideset{}{'}{\sum}_{k=1}^{[q/2]}(-1)^{k}\cos (\pi k/q)\cos (2\pi
k\alpha _{0})f_{3/2}\left( 2mr\sin (\pi k/q)\right) \right] .  \label{FCcsM}
\end{eqnarray}%
Note that the function $f_{3/2}\left( x\right) $ is expressed in terms of
elementary functions:%
\begin{equation}
f_{3/2}\left( x\right) =\sqrt{\frac{\pi }{2}}\frac{x+1}{x^{3}}e^{-x}.
\label{f32}
\end{equation}%
For a massless field the condensate (\ref{FCcsM}) vanishes. This result is
also seen from (\ref{FCm0}) taking the limit $a\rightarrow \infty $ with
fixed $y$. Hence, the generation of the FC for a massless field is purely
gravitational effect.

At small proper distances from the string compared with the curvature radius
we have $\rho \ll 1$ and for a massive field one gets 
\begin{equation}
\langle \bar{\psi}\psi \rangle _{\mathrm{cs}}\approx -\frac{mh_{3}(q,\alpha
_{0})}{8\pi ^{2}(ar/w)^{3}}\ .  \label{Fccsrsm}
\end{equation}%
As seen, in this case the FC diverges on the string like $1/r_{p}^{3}$. The
leading term (\ref{Fccsrsm}) coincides with that for the string in the
Minkowski bulk, given by (\ref{FCcsM}), replacing the Minkowskian distance $%
r $ by the proper distance $r_{p}=ar/w$ in the AdS bulk. This shows that for
a massive field the effects of gravity near the string are weak. At large
distances from the string's core, $\rho \gg 1$, with $w$ fixed, we use the
asymptotic 
\begin{equation}
{\mathcal{Z}}_{ma}(u)\approx \frac{\sqrt{\pi }(ma+1/2)(ma+3/2)}{%
2^{ma}u^{ma+5/2}}\ ,  \label{asymp1}
\end{equation}%
for $u\gg 1$. Pluging this into (\ref{FC-cs}), to the leading order, we
obtain 
\begin{equation}
\langle \bar{\psi}\psi \rangle _{\mathrm{cs}}\approx -\frac{\left(
ma+1/2\right) (ma+3/2)}{2^{2ma+2}\pi ^{2}a^{4}(r/w)^{2ma+5}}%
h_{2ma+5}(q,\alpha _{0})\ .  \label{larger}
\end{equation}%
For a massless field this result could also be directly obtained from (\ref%
{FCm0}). We see that the decay of the FC at large distance from the string
is power-law for both massless and massive fields. Note that, as it is seen
from (\ref{FCcsM}), in the Minkowski bulk at large distances, $mr\gg 1$, the
FC decays exponentially.

By taking into account that the FC depends on the coordinates $r$ and $w$ in
the form of the ratio $r/w$, from the asymptotic expressions given above for 
$\rho \ll 1$ and $\rho \gg 1$ we can obtain the behavior of the condensate
near the AdS boundary and horizon. For points near the boundary, assuming
that $w\ll r$, the leading term in the corresponding asymptotic expansion is
given by (\ref{larger}) and the condensate $\langle \bar{\psi}\psi \rangle _{%
\mathrm{cs}}$ vanishes on the AdS boundary as $w^{2ma+5}$. Near the horizon
one has $w\gg r$ and for a massive field, in accordance with (\ref{Fccsrsm}%
), the contribution of the cosmic string in the FC diverges on the horizon
like $w^{3}$. For a massless field and under the condition (\ref{cond2}) the
divergence is weaker, $\langle \bar{\psi}\psi \rangle _{\mathrm{cs}}\propto
w^{1+\left( 2|\alpha _{0}|-1\right) q}$. In the range of parameters $%
2|\alpha _{0}|<1-1/q$ the condensate $\langle \bar{\psi}\psi \rangle _{%
\mathrm{cs}}$ is finite on the horizon with the limiting value given by the
right-hand side of (\ref{FCm0r0}).

\begin{figure}[tbph]
\begin{center}
\begin{tabular}{cc}
\epsfig{figure=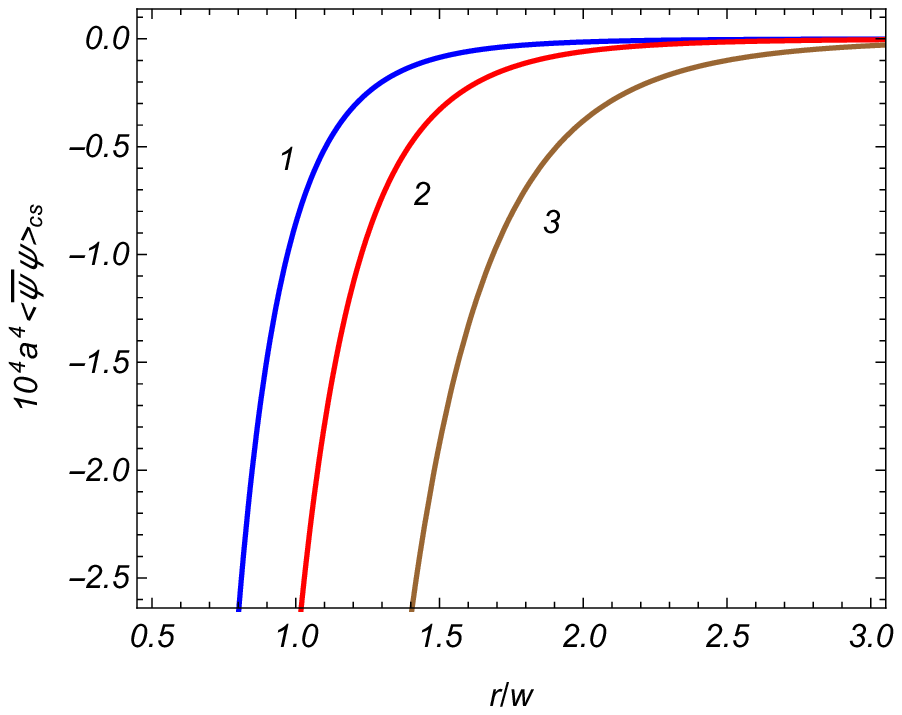,width=7.cm,height=5.5cm} & \quad %
\epsfig{figure=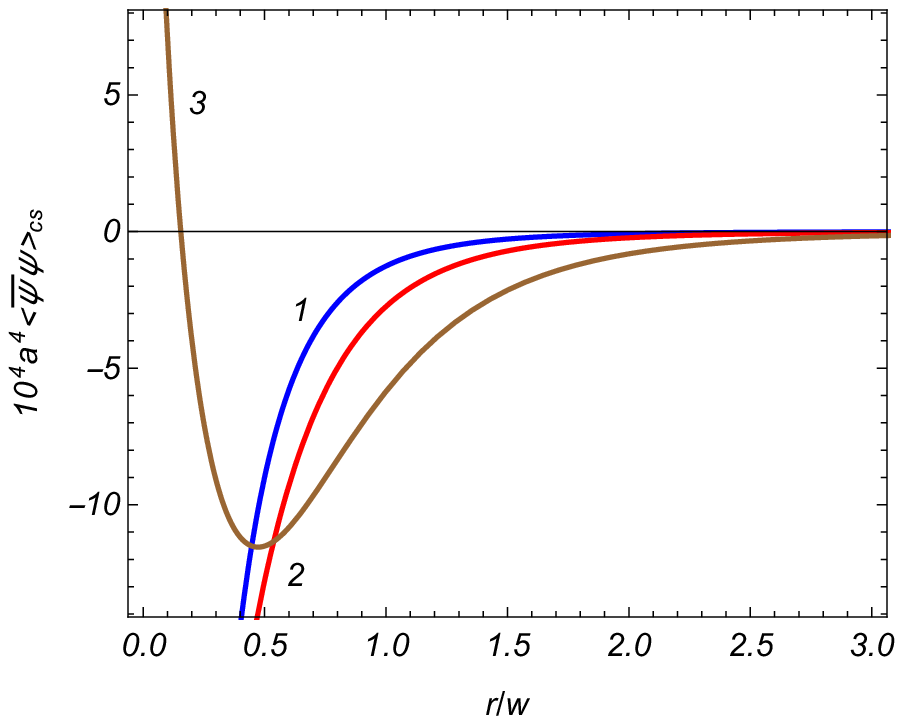,width=7.cm,height=5.5cm}%
\end{tabular}%
\end{center}
\caption{The FC $\langle \bar{\protect\psi}\protect\psi \rangle _{\mathrm{cs}%
}$ as a function of the distance from the string for massive (left panel)
and massless (right panel) fields. The graphs are plotted for $\protect%
\alpha _{0}=0.3$ and for different values of $q$ (numbers near the curves).
On the left panel we have taken $ma=1$.}
\label{fig1}
\end{figure}

In Fig. \ref{fig1} we have plotted the FC in the geometry where the $z$%
-direction is not compactified as a function of the ratio $r/w$. The latter
is the proper distance from the string in units of the curvature radius $a$,
measured by an observer with the coordinate $w$. The graphs are plotted for $%
\alpha _{0}=0.3$ and the numbers near the curves are the values of the
parameter $q$. The left panel corresponds to a massive fermionic field with $%
ma=1$ and the right panel presents the results for a massless field. As it
has been already clarified by the asymptotic analysis, for a massive field
the condensate $\langle \bar{\psi}\psi \rangle _{\mathrm{cs}}$ diverges on
the string like $(w/r)^{3}$. For a massless field, $\langle \bar{\psi}\psi
\rangle _{\mathrm{cs}}$ diverges as $\left( w/r\right) ^{1+\left( 2|\alpha
_{0}|-1\right) q}$ under the condition (\ref{cond2}) (the curves with $q=1,2$
on the right panel) and takes finite value (\ref{FCm0r0}) on the string
under the condition (\ref{cond1}) (the curve $q=3$ on the right panel). Note
that for the case $q=3$ on the right panel $\langle \bar{\psi}\psi \rangle _{%
\mathrm{cs}}|_{r=0}\approx 0.0095/a^{4}$.

\begin{figure}[tbph]
\begin{center}
\begin{tabular}{cc}
\epsfig{figure=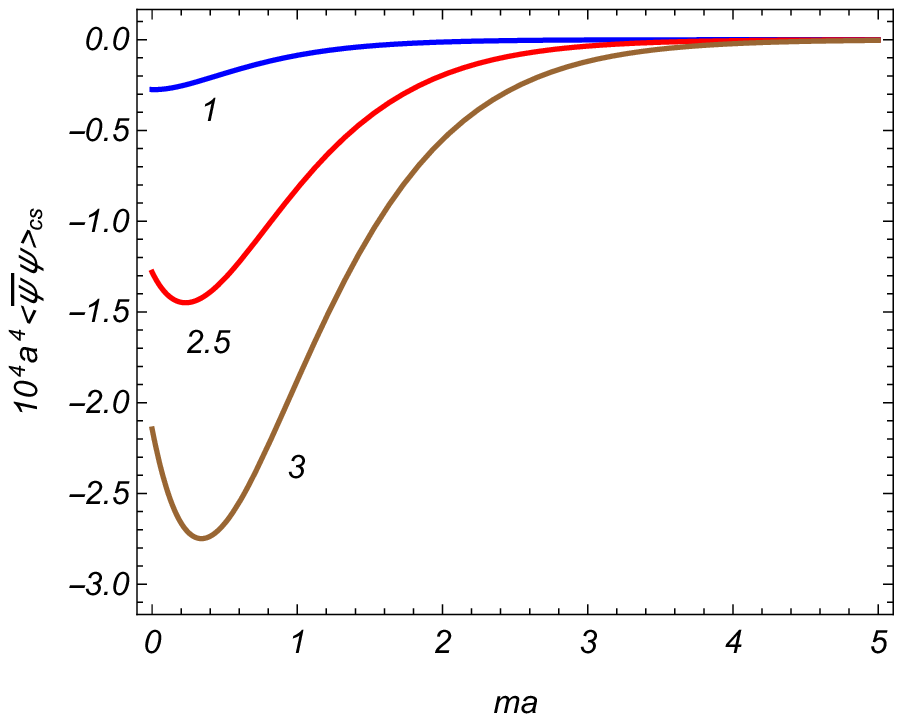,width=7.cm,height=5.5cm} & \quad %
\epsfig{figure=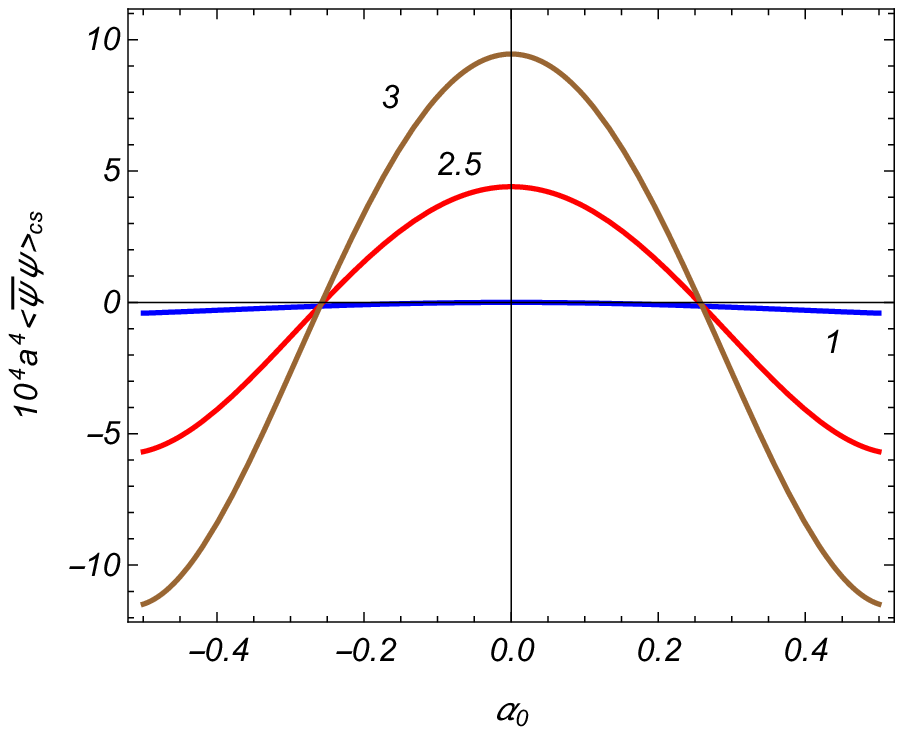,width=7.cm,height=5.5cm}%
\end{tabular}%
\end{center}
\caption{The condensate $\langle \bar{\protect\psi}\protect\psi \rangle _{%
\mathrm{cs}}$ as a function of the mass (left panel) and of the parameter $%
\protect\alpha _{0}$ (right panel) for fixed $r/w=1.5$ and for different
values of $q$ (numbers near curves). For the left panel we have taken $%
\protect\alpha _{0}=0.3$ and for the right panel $ma=0.5$.}
\label{fig2}
\end{figure}

The left panel of Fig. \ref{fig2} presents the dependence of the FC $\langle 
\bar{\psi}\psi \rangle _{\mathrm{cs}}$ on the mass of the field, in units of 
$1/a$, for $r/w=1.5$ and $\alpha _{0}=0.3$. The right panel in Fig. \ref%
{fig2} displays the dependence of the FC $\langle \bar{\psi}\psi \rangle _{%
\mathrm{cs}}$ on $\alpha _{0}$ for fixed values of $ma=0.5$ and $r/w=1.5$.
On both panels the numbers near the curves correspond to the values of the
parameter $q$.

We recall that the consideration above was presented in terms of the fields $%
\psi _{(s)}^{\prime }$ with the Lagrangian density (\ref{Lsp}) and the
formulas given above are for the FC $\langle \bar{\psi}_{(s)}^{\prime }\psi
_{(s)}^{\prime }\rangle $ with $s=+1$. As it has been shown above 
\begin{equation}
\langle \bar{\psi}_{(-1)}^{\prime }\psi _{(-1)}^{\prime }\rangle =-\langle 
\bar{\psi}_{(+1)}^{\prime }\psi _{(+1)}^{\prime }\rangle .  \label{FCpr}
\end{equation}%
Having the results for $\langle \bar{\psi}_{(s)}^{\prime }\psi
_{(s)}^{\prime }\rangle $ we can obtain the corresponding quantity in terms
of the initial fields $\psi _{(s)}$ with the Lagrangian density (\ref{Ls}).
By using the relation between the fields $\psi _{(s)}$ and $\psi
_{(s)}^{\prime }$ one can see that%
\begin{equation}
\langle \bar{\psi}_{(s)}\psi _{(s)}\rangle =s\langle \bar{\psi}%
_{(s)}^{\prime }\psi _{(s)}^{\prime }\rangle .  \label{FCssp}
\end{equation}%
From here we conclude that the FC $\langle \bar{\psi}_{(s)}\psi
_{(s)}\rangle $ is the same for two irreducible representations $s=+1$ and $%
s=-1$. For (1+4)-dimensional spacetime the mass term in the Lagrangian
density (\ref{Ls}) is not invariant under the parity transformation ($P$)
and charge conjugation ($C$). Invariant massive fermionic models can be
constructed considering a system of two 4-component fields $\psi _{(+1)}$
and $\psi _{(-1)}$ with the Lagrangian density $L=\sum_{s=\pm 1}L_{(s)}$. In
those models the total FC induced by cosmic string in the uncompactified
geometry is obtained summing the separate contributions, $\sum_{s=\pm
1}\langle \bar{\psi}_{(s)}\psi _{(s)}\rangle _{\mathrm{cs}}$. These
contributions coincide for the fields with $s=+1$ and $s=-1$ and are given
by the expressions for $\langle \bar{\psi}\psi \rangle _{\mathrm{cs}}$
presented above.

\section{Topological effects of compactification}

\label{sec:Comp}

In this section we consider the FC in the geometry with compactified $z$%
-direction. The scalar and fermionic vacua polarizations around a
compactified cosmic string in the Minkowksi bulk have been investigated in 
\cite{Beze12co,Beze13co,Bell14co}. The mode sum for the FC is still given by
(\ref{FC}) where now the mode functions are expressed as (\ref{wfunc}) and
the collective summation is understood as 
\begin{equation}
\sum_{\sigma }=\int_{0}^{\infty }d\lambda \int_{0}^{\infty
}dp\sum_{j}\sum_{l=-\infty }^{\infty }\sum_{\eta =\pm 1}\,.  \label{Sumsigc}
\end{equation}%
Substituting the mode functions, in a way similar to (\ref{AdScs}), we find
the representation%
\begin{eqnarray}
\langle \bar{\psi}\psi \rangle &=&-\frac{sqw^{5}}{2\pi a^{4}L}%
\int_{0}^{\infty }d\lambda \int_{0}^{\infty }dp\sum_{j}\sum_{l=-\infty
}^{\infty }\frac{\lambda p^{2}}{\sqrt{\lambda ^{2}+p^{2}+\tilde{k}_{l}^{2}}}
\notag \\
&&\times J_{\nu _{1}}(pw)J_{\nu _{2}}(pw)\left[ J_{\beta _{j}}^{2}(\lambda
r)+J_{\beta _{j}+\epsilon _{j}}^{2}(\lambda r)\right] \ .  \label{FC3}
\end{eqnarray}%
Again, the FC has opposite signs in the cases $s=+1$ and $s=-1$ and we
continue the investigation for $s=+1$ and, hence, in the discussion below $%
\nu _{1}=ma-1/2$, $\nu _{2}=ma+1/2$.

In order to separate the contribution in the FC induced by the
compactification of the $z$-direction, for the summation over $l$ we use the
Abel-Plana-type formula \cite{Bell10AP} 
\begin{equation}
\sum_{l=-\infty }^{\infty }f(|l+\tilde{\beta}|)=2\int_{0}^{\infty
}du\,f(u)+i\int_{0}^{\infty }du\sum_{n=\pm 1}\frac{f(iu)-f(-iu)}{e^{2\pi
(u+in\tilde{\beta})}-1},  \label{AP}
\end{equation}%
with the function $f(u)=[\lambda ^{2}+p^{2}+(2\pi u/L)^{2}]^{-1/2}$.
Comparing with (\ref{AdScs}), we see that the contribution to the FC\ coming
from the first integral in (\ref{AP}) coincides with the FC in the
uncompactified geometry. In this way the following decomposition is obtained:%
\begin{equation}
\langle \bar{\psi}\psi \rangle =\langle \bar{\psi}\psi \rangle _{\mathrm{cs}%
}^{\mathrm{AdS}}+\langle \bar{\psi}\psi \rangle _{\mathrm{c}}\,,
\label{FC_components}
\end{equation}%
where the second term in the right-hand side comes from the second integral
in (\ref{AP}) and contains the effects induced by the compactification. For
that contribution one gets%
\begin{eqnarray}
\langle \bar{\psi}\psi \rangle _{\mathrm{c}} &=&-\frac{qw^{5}}{2\pi ^{2}a^{4}%
}\int_{0}^{\infty }d\lambda \,\lambda \int_{0}^{\infty
}dp\,p^{2}\sum_{j}[J_{\beta _{j}}^{2}(\lambda r)+J_{\beta _{j}+\epsilon
_{j}}^{2}(\lambda r)]  \notag \\
&&\times J_{\nu _{1}}(pw)J_{\nu _{2}}(pw)\int_{\sqrt{\lambda ^{2}+p^{2}}%
}^{\infty }dx\sum_{n=\pm 1}\frac{\left( x^{2}-\lambda ^{2}-p^{2}\right)
^{-1/2}}{e^{Lx+2\pi in\tilde{\beta}}-1}\ ,  \label{FC-compact}
\end{eqnarray}%
where a new integration variable $x=2\pi u/L$ is introduced. Note that for $%
r>0$ the topological part $\langle \bar{\psi}\psi \rangle _{\mathrm{c}}$ is
finite and implicit regularization assumed before can be removed. The
renormalization is required for the part $\langle \bar{\psi}\psi \rangle _{%
\mathrm{cs}}^{\mathrm{AdS}}$ only and that has been discussed in the
previous section. The physical reason for the finiteness of $\langle \bar{%
\psi}\psi \rangle _{\mathrm{c}}$ is that the structure of divergences is
completely determined by the local geometry and the latter is not changed by
the compactification under consideration.

For the further transformation of the condensate $\langle \bar{\psi}\psi
\rangle _{\mathrm{c}}$ we use the expansion $(e^{u}-1)^{-1}=\sum_{l=1}^{%
\infty }e^{-lu}$ in the integral over $x$. The integrals for a given $l$ are
expressed in terms of the Macdonald function and one obtains 
\begin{eqnarray}
\langle \bar{\psi}\psi \rangle _{\mathrm{c}} &=&-\frac{qw^{5}}{\pi ^{2}a^{4}}%
\sum_{l=1}^{\infty }\cos (2\pi l\tilde{\beta})\int_{0}^{\infty
}dp\,p^{2}J_{\nu _{1}}(pw)J_{\nu _{2}}(pw)  \notag \\
&&\times \sum_{j}\int_{0}^{\infty }d\lambda \,\lambda \lbrack J_{\beta
_{j}}^{2}(\lambda r)+J_{\beta _{j}+\epsilon _{j}}^{2}(\lambda r)]K_{0}(lL%
\sqrt{\lambda ^{2}+p^{2}})\ .  \label{FC_C}
\end{eqnarray}%
Using the integral representation 
\begin{equation}
K_{\nu }(z)=\frac{1}{2}\left( \frac{z}{2}\right) ^{\nu }\int_{0}^{\infty
}dt\,\frac{e^{-t-z^{2}/4t}}{t^{\nu +1}}\ ,  \label{Macdonald}
\end{equation}%
the integrals over $\lambda $ and $p$ in (\ref{FC_C}) are expressed through
the modified Bessel function $I_{\nu }\left( x\right) $. With the notation (%
\ref{Jcal}), the compactification part in the FC is presented as 
\begin{eqnarray}
\langle \bar{\psi}\psi \rangle _{\mathrm{c}} &=&-\frac{q}{2\pi ^{2}a^{4}}%
\sum_{l=1}^{\infty }\cos (2\pi l\tilde{\beta})\int_{0}^{\infty
}dx\,x^{2}e^{-[1+\rho ^{2}+(lL)^{2}/2w^{2}]x}  \notag \\
&&\times \left[ I_{\nu _{1}}\left( x\right) -I_{\nu _{2}}\left( x\right) %
\right] {\mathcal{J}}(q,\alpha _{0},x\rho ^{2})\ .  \label{FCc}
\end{eqnarray}%
In this expression we have introduced a new variable, $x=2tw^{2}/(lL)^{2}$.

Using the representation for the function ${\mathcal{J}}(q,\alpha _{0},x)$
given in (\ref{J-function}), we obtain 
\begin{eqnarray}
\langle \bar{\psi}\psi \rangle _{\mathrm{c}} &=&-\frac{2}{\pi ^{2}a^{4}}%
\sum_{l=1}^{\infty }\cos (2\pi l\tilde{\beta})\int_{0}^{\infty
}dx\,x^{2}e^{-[1+(lL)^{2}/2w^{2}]x}\left[ I_{\nu _{1}}(x)-I_{\nu _{2}}(x)%
\right]  \notag \\
&&\times \left[ \sideset{}{'_*}{\sum}_{k=0}^{[q/2]}(-1)^{k}\cos (\pi
k/q)\cos (2\pi k\alpha _{0})e^{-2x\rho ^{2}\sin ^{2}(\pi k/q)}\right.  \notag
\\
&&\left. +\frac{q}{\pi }\int_{0}^{\infty }du\frac{h(q,\alpha _{0},2u)\sinh u%
}{\cosh (2qu)-\cos (q\pi )}e^{-2x\rho ^{2}\cosh ^{2}u}\right] \ ,
\label{compc}
\end{eqnarray}%
where the asterisk sign in the summation over $k$ in (\ref{compc}) indicates
that the term $k=0$ must be divided by $2$. The integral over $x$ is
evaluated by using the formula (\ref{function_Q}) and we get the final
expression 
\begin{eqnarray}
\langle \bar{\psi}\psi \rangle _{\mathrm{c}} &=&-\frac{2^{3/2}}{\pi
^{5/2}a^{4}}\sum_{l=1}^{\infty }\cos (2\pi l\tilde{\beta})\Biggl[%
\sideset{}{'_*}{\sum}_{k=0}^{[q/2]}(-1)^{k}\cos (\pi k/q)\cos (2\pi k\alpha
_{0}){\mathcal{Z}}_{ma}(u_{lk})  \notag \\
&&+\frac{q}{\pi }\int_{0}^{\infty }du\frac{h(q,\alpha _{0},2u)\sinh u}{\cosh
(2qu)-\cos (q\pi )}{\mathcal{Z}}_{ma}(u_{lu})\Biggr]\ ,  \label{FC-c}
\end{eqnarray}%
where we have defined new variables 
\begin{eqnarray}
&&u_{lk}=1+2\rho ^{2}\sin ^{2}(\pi k/q)+\frac{(lL)^{2}}{2w^{2}}\ ,  \notag \\
&&u_{lu}=1+2\rho ^{2}\cosh ^{2}u+\frac{(lL)^{2}}{2w^{2}}\ ,  \label{args-c}
\end{eqnarray}%
and the function ${\mathcal{Z}}_{ma}(u)$ is given by (\ref{Z_function}).

The compactification part (\ref{FC-c}) depends on $r$, $L$, $w$ in the form
of the ratios $r/w$ and $L/w$. Again, that is a consequence of the maximal
symmetry of the AdS spacetime. Note that $L/w$ is the proper length of the
compact dimension measured by an observer with a given value of the
coordinate $w$. The $k=0$ term in (\ref{FC-c}) is presented as 
\begin{equation}
\langle \bar{\psi}\psi \rangle _{\mathrm{c}}^{(0)}=-\frac{\sqrt{2}}{\pi
^{5/2}a^{4}}\sum_{l=1}^{\infty }\cos (2\pi l\tilde{\beta})\mathcal{Z}%
_{ma}\left( 1+\frac{l^{2}L^{2}}{2w^{2}}\right) .  \label{FCk0}
\end{equation}%
For $q=1$ and $\alpha _{0}=0$ this part survives only in (\ref{FC-c}) and,
hence, it corresponds to the contribution in the FC induced by the
compactification of the $z$-direction in the AdS spacetime when the cosmic
string is absent. The remaining part in (\ref{FC-c}), corresponding to the
difference $\langle \bar{\psi}\psi \rangle _{\mathrm{c}}-\langle \bar{\psi}%
\psi \rangle _{\mathrm{c}}^{(0)}$, is induced by the planar angle deficit
and magnetic flux. In the range (\ref{cond1}) of the parameters the
topological contribution (\ref{FC-c}) is finite on the string:%
\begin{equation}
\langle \bar{\psi}\psi \rangle _{\mathrm{c}}|_{r=0}=\left[ 1+2h_{0}(q,\alpha
_{0})\right] \langle \bar{\psi}\psi \rangle _{\mathrm{c}}^{(0)}.
\label{FCcr0}
\end{equation}%
For $2|\alpha _{0}|>1-1/q$ the FC $\langle \bar{\psi}\psi \rangle _{\mathrm{c%
}}$ diverges on the string like $1/\rho ^{1+\left( 2|\alpha _{0}|-1\right)
q} $. This can be seen in the way similar to that we have used for (\ref%
{FCcsNear}).

The topological part in the FC, induced by the cosmic string and by the
compactification of the $z$-direction, is given by $\langle \bar{\psi}\psi
\rangle _{\mathrm{t}}=$ $\langle \bar{\psi}\psi \rangle -\langle \bar{\psi}%
\psi \rangle _{\mathrm{AdS}}$ and it is presented as the sum%
\begin{equation}
\langle \bar{\psi}\psi \rangle _{\mathrm{t}}=\langle \bar{\psi}\psi \rangle
_{\mathrm{cs}}+\langle \bar{\psi}\psi \rangle _{\mathrm{c}}.  \label{FCt}
\end{equation}%
By taking into account the formulas (\ref{FC-cs}) and (\ref{FC-c}), the
corresponding expression reads%
\begin{eqnarray}
\langle \bar{\psi}\psi \rangle _{\mathrm{t}} &=&\langle \bar{\psi}\psi
\rangle _{\mathrm{c}}^{(0)}-\frac{2^{3/2}}{\pi ^{5/2}a^{4}}%
\sideset{}{'}{\sum}_{l=0}^{\infty }\cos (2\pi l\tilde{\beta})\Biggl[%
\sideset{}{'}{\sum}_{k=1}^{[q/2]}(-1)^{k}\cos (\pi k/q)\cos (2\pi k\alpha
_{0}){\mathcal{Z}}_{ma}(u_{lk})  \notag \\
&&+\frac{q}{\pi }\int_{0}^{\infty }du\frac{h(q,\alpha _{0},2u)\sinh u}{\cosh
(2qu)-\cos (q\pi )}{\mathcal{Z}}_{ma}(u_{lu})\Biggr]\ ,  \label{FCt1}
\end{eqnarray}%
where the prime on the summation sign over $l$ means that the term with $l=0$
should be taken with coefficient 1/2.

The expression (\ref{FC-c}) is further simplified for a massless field. By
using (\ref{massless-Z}) we get 
\begin{eqnarray}
\langle \bar{\psi}\psi \rangle _{\mathrm{c}} &=&-\frac{3}{8\pi ^{2}a^{4}}%
\sum_{l=1}^{\infty }\cos (2\pi l\tilde{\beta})\Biggl[\sideset{}{'_*}{\sum}%
_{k=0}^{[q/2]}\frac{(-1)^{k}\cos (\pi k/q)\cos (2\pi k\alpha _{0})}{\left[
1+\rho ^{2}\sin ^{2}(\pi k/q)+(lL/w)^{2}/4\right] ^{5/2}}  \notag \\
&&+\frac{q}{\pi }\int_{0}^{\infty }du\frac{\sinh u}{\cosh (2qu)-\cos (q\pi )}%
\frac{h(q,\alpha _{0},2u)}{\left[ 1+\rho ^{2}\cosh ^{2}u+(lL/w)^{2}/4\right]
^{5/2}}\Biggr]\ .  \label{FCcm0}
\end{eqnarray}%
Similar to the straight cosmic string part (\ref{FCm0}), under the condition
(\ref{cond1}) the compactification contribution (\ref{FCcm0}) is finite on
the string:%
\begin{equation}
\langle \bar{\psi}\psi \rangle _{\mathrm{c}}|_{r=0}=-3\frac{%
1+2h_{0}(q,\alpha _{0})}{16\pi ^{2}a^{4}}\sum_{l=1}^{\infty }\frac{\cos
(2\pi l\tilde{\beta})}{\left[ 1+(lL/2w)^{2}\right] ^{5/2}}\ ,
\label{FCcm0r0}
\end{equation}%
and diverges as $1/\rho ^{1+\left( 2|\alpha _{0}|-1\right) q}$ for $2|\alpha
_{0}|>1-1/q$\thinspace .

In the special case $q=1$, corresponding to the zero planar angle deficit,
the contribution in the FC induced by the compactification of the $z$%
-coordinate is presented as%
\begin{eqnarray}
\langle \bar{\psi}\psi \rangle _{\mathrm{c}} &=&-\frac{2^{3/2}\sin (\pi
\alpha _{0})}{\pi ^{7/2}a^{4}}\sum_{l=1}^{\infty }\cos (2\pi l\tilde{\beta}%
)\int_{0}^{\infty }dx\,\tanh x  \notag \\
&&\times \sinh (2\alpha _{0}x){\mathcal{Z}}_{ma}(1+2\rho ^{2}\cosh
^{2}x+(lL/w)^{2}/2)\ .  \label{FCcq1}
\end{eqnarray}%
The topological part of the FC is given by $\langle \bar{\psi}\psi \rangle _{%
\mathrm{t}}=\langle \bar{\psi}\psi \rangle _{\mathrm{mf}}+\langle \bar{\psi}%
\psi \rangle _{\mathrm{c}}$, where the FC induced by the magnetic flux in
the uncompactified geometry is expressed as (\ref{FCmf}). For $|\alpha
_{0}|>0$, the condensate (\ref{FCcq1}) diverges on the location of the
magnetic flux like $1/\rho ^{2|\alpha _{0}|}$. Other asymptotics are
obtained from the corresponding expressions for general $q$ with the
function $h_{n}(1,\alpha _{0})$ from (\ref{hn1}).

In the Minkowskian limit, corresponding to $a\rightarrow \infty $ with fixed 
$y$, by using the result (\ref{AsMink}), we obtain%
\begin{eqnarray}
\langle \bar{\psi}\psi \rangle _{\mathrm{c}}^{\mathrm{(M)}} &=&-\frac{2^{3/2}%
}{\pi ^{5/2}}m^{4}\sum_{l=1}^{\infty }\cos (2\pi l\tilde{\beta})  \notag \\
&&\times \left[ \sideset{}{'_*}{\sum}_{k=0}^{[q/2]}(-1)^{k}\cos (\pi
k/q)\cos (2\pi k\alpha _{0})f_{3/2}\left( 2m\sqrt{r^{2}\sin ^{2}(\pi
k/q)+l^{2}L^{2}/4}\right) \right.  \notag \\
&&\left. +\frac{q}{\pi }\int_{0}^{\infty }du\frac{h(q,\alpha _{0},2u)\sinh u%
}{\cosh (2qu)-\cos (q\pi )}f_{3/2}\left( 2m\sqrt{r^{2}\cosh
^{2}u+l^{2}L^{2}/4}\right) \right] \ ,  \label{FCcM}
\end{eqnarray}%
where the function $f_{3/2}(x)$ is given by (\ref{f32}). For a massless
field the topological contribution vanishes. Again we see that the nonzero
FC for a massless field on AdS bulk is a gravitationally induced effect.
Similar to the case of the AdS bulk, the $k=0$ term in (\ref{FCcM}), given
by 
\begin{equation}
\langle \bar{\psi}\psi \rangle _{\mathrm{c}}^{(0)\mathrm{(M)}}=-\frac{\sqrt{2%
}}{\pi ^{5/2}}m^{4}\sum_{l=1}^{\infty }\cos (2\pi l\tilde{\beta}%
)f_{3/2}\left( mlL\right) ,  \label{FCcM0}
\end{equation}%
is the FC in (1+4)-dimensional Minkowski spacetime with compactified $z$%
-direction in the absence of cosmic string. It does not depend on the radial
coordinate. The part $\langle \bar{\psi}\psi \rangle _{\mathrm{c}}^{\mathrm{%
(M)}}-\langle \bar{\psi}\psi \rangle _{\mathrm{c}}^{(0)\mathrm{(M)}}$ is
induced by the presence of cosmic string and for a massive field it
exponentially decays at large distances from the string.

In order to see the asymptotic behavior of the FC on the AdS bulk at large
distances from the string, $\rho \gg 1$, we use in (\ref{FC-c}) the
asymptotic formula (\ref{asymp1}). Then, by taking into account that for the
leading term of the series over $l$ one has%
\begin{equation}
\sum_{l=1}^{\infty }\frac{\cos (2\pi l\tilde{\beta})}{\left(
b^{2}+l^{2}x^{2}\right) ^{ma+5/2}}\approx -\frac{1}{2b^{2ma+5}},\;x\ll 1,
\label{seras}
\end{equation}%
the FC (\ref{FC-c}) is estimated as 
\begin{equation}
\langle \bar{\psi}\psi \rangle _{\mathrm{c}}\approx \langle \bar{\psi}\psi
\rangle _{\mathrm{c}}^{(0)}+\frac{(ma+1/2)(ma+3/2)}{2^{2ma+2}\pi
^{2}a^{4}(r/w)^{2ma+5}}h_{2ma+5}(q,\alpha _{0}).  \label{FCclarger}
\end{equation}%
We see that at large distances from the string the contribution in $\langle 
\bar{\psi}\psi \rangle _{\mathrm{c}}$ coming from the magnetic flux and from
the planar angle deficit decays as $(w/r)^{2ma+5}$. Comparing (\ref%
{FCclarger}) with the corresponding expansion for the straight cosmic string
part, given by (\ref{larger}), we see that the decay of the corresponding
contribution in the total FC (given by $\langle \bar{\psi}\psi \rangle
-\langle \bar{\psi}\psi \rangle _{\mathrm{AdS}}-\langle \bar{\psi}\psi
\rangle _{\mathrm{c}}^{(0)}$) is stronger than in separate terms $\langle 
\bar{\psi}\psi \rangle _{\mathrm{cs}}$ and $\langle \bar{\psi}\psi \rangle _{%
\mathrm{c}}$. As seen from (\ref{FCclarger}), the contribution in the FC $%
\langle \bar{\psi}\psi \rangle _{\mathrm{c}}$ induced by the cosmic string,
as a function of the proper distance from the string, decays according to a
power law. This behavior is in contrast to the exponential decay for cosmic
string in the Minkowski bulk.

\begin{figure}[tbph]
\begin{center}
\begin{tabular}{cc}
\epsfig{figure=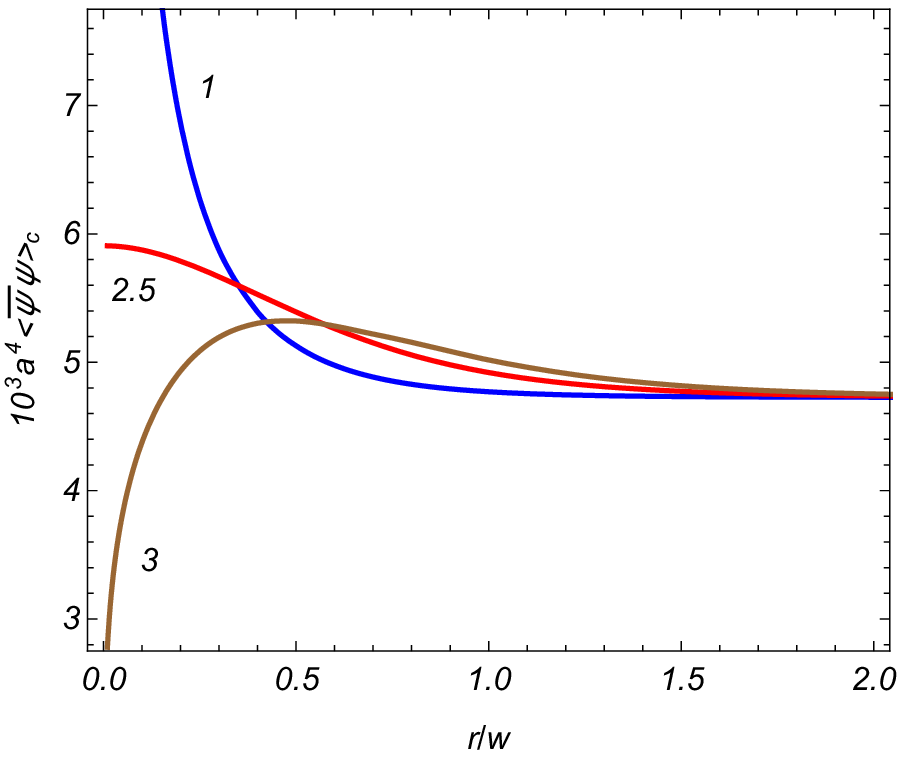,width=7.cm,height=5.5cm} & \quad %
\epsfig{figure=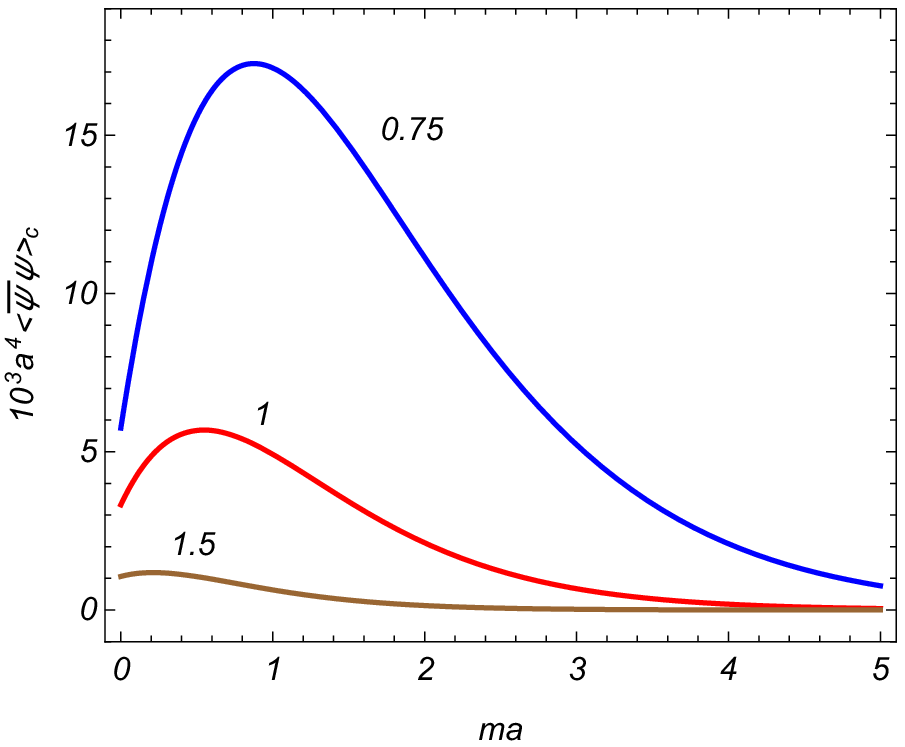,width=7.cm,height=5.5cm}%
\end{tabular}%
\end{center}
\caption{The part in the FC induced by compactification as a function of the
radial distance from the string (left panel) and as a function of the mass
(right panel). For both panels $\protect\alpha _{0}=0.3$, $\tilde{\protect%
\beta}=0.25$. The numbers near the curves are the values of the parameter $q$
for the left panel and of the ratio $L/w$ for the right panel. The graphs
are plotted for $ma=1$, $L/w=1$ and for $q=2.5$, $r/w=1$ on the left and
right panels, respectively.}
\label{fig3}
\end{figure}

In Fig. \ref{fig3} the contribution in the FC, induced by the
compactification, is plotted versus the radial distance from the string
(left panel) and the mass of the field (right panel). For the graphs on the
left panel we have taken $ma=1$, $L/w=1$, $\alpha _{0}=0.3$, $\tilde{\beta}%
=0.25$ and the numbers near the curves correspond to the values of $q$. For
the right panel the parameters are fixed as $q=2.5$, $r/w=1$, $\alpha
_{0}=0.3$, $\tilde{\beta}=0.25$ and the numbers near the curves are the
values of the ratio $L/w$. As it has been explained above, under the
condition (\ref{cond1}) the FC $\langle \bar{\psi}\psi \rangle _{\mathrm{c}}$
is finite on the string. This corresponds to the curve with $q=3$ on the
left panel of Fig. \ref{fig3}. In the range of the parameters corresponding
to $2|\alpha _{0}|>1-1/q$, the condensate $\langle \bar{\psi}\psi \rangle _{%
\mathrm{c}}$ diverges on the string like $1/r^{1+\left( 2|\alpha
_{0}|-1\right) q}$ and this situation is exemplified by the curve with $q=1$
on the left panel. For the curve with $q=2.5$ one has $2|\alpha _{0}|=1-1/q$
and the topological part is finite on the string. At large distances from
the string the condensate tends to the limiting value $\langle \bar{\psi}%
\psi \rangle _{\mathrm{c}}^{(0)}$ that does not depend on $q$ and $\alpha
_{0}$. An interesting feature in the dependence on the mass is that the FC
takes its maximum for some intermediate value of the mass. Of course, we
could expect the suppression of the condensate for large masses.

\begin{figure}[tbph]
\begin{center}
\begin{tabular}{cc}
\epsfig{figure=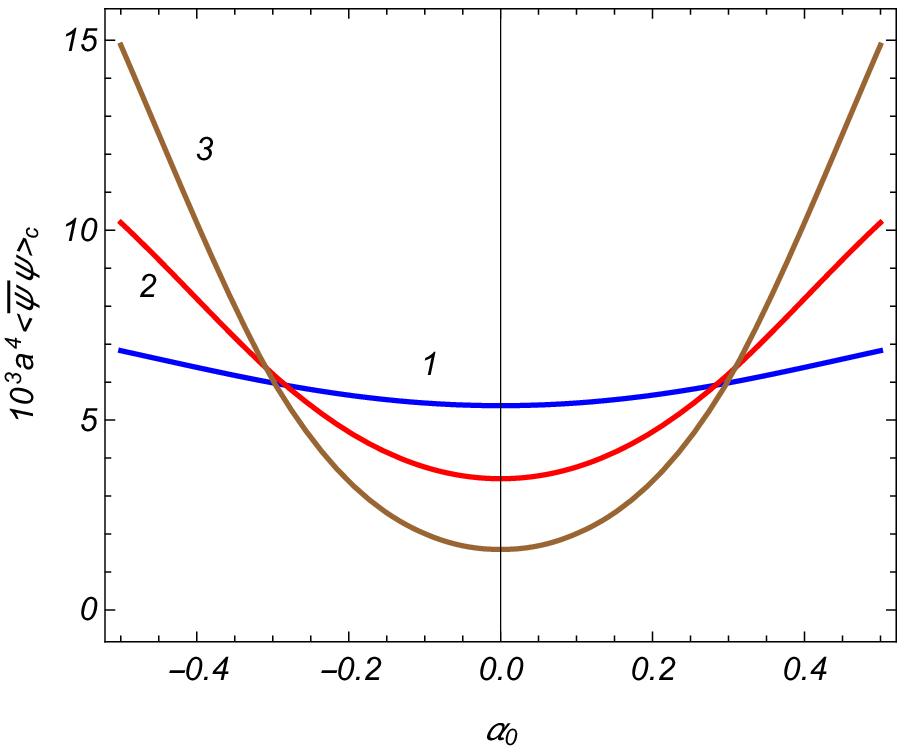,width=7.cm,height=5.5cm} & \quad %
\epsfig{figure=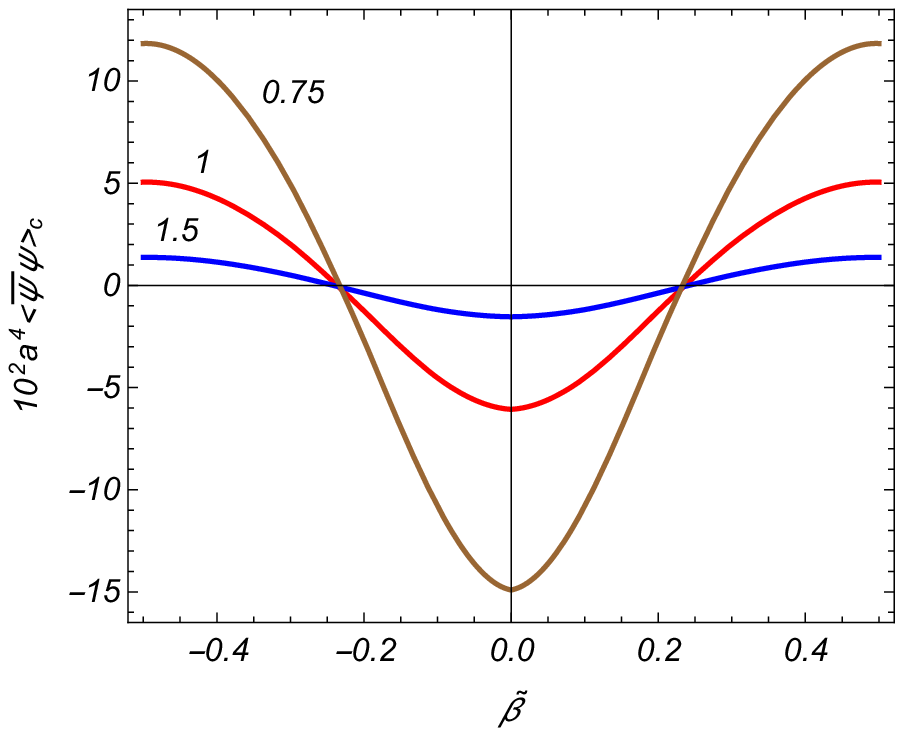,width=7.cm,height=5.5cm}%
\end{tabular}%
\end{center}
\caption{The condensate $\langle \bar{\protect\psi}\protect\psi \rangle _{%
\mathrm{c}}$ versus the parameters $\protect\alpha _{0}$ and $\tilde{\protect%
\beta}$. For both panels $ma=0.5$, $r/w=0.5$. The numbers near the curves
are the values of the parameter $q$ for the left panel and of the ratio $L/w$
for the right panel. The graphs are plotted for $L/w=1$, $\tilde{\protect%
\beta}=0.25$ and for $q=2$, $\protect\alpha _{0}=0.2$ on the left and right
panels, respectively.}
\label{fig4}
\end{figure}

Figure \ref{fig4} presents the part $\langle \bar{\psi}\psi \rangle _{%
\mathrm{c}}$ in the FC as a function of $\alpha _{0}$ (left panel) and of $%
\tilde{\beta}$ (right panel). The left panel is plotted for $ma=0.5$, $%
r/w=0.5$, $L/w=1$, $\tilde{\beta}=0.25$ and the numbers near the curves
correspond to the values of the parameter $q$. For the right panel we have
taken $q=2$, $ma=0.5$, $r/w=0.5$, $\alpha _{0}=0.2$ and the numbers near the
curves are the values of the ratio $L/w$. Considered as a function of the
parameter $\tilde{\beta}$, the FC and its first derivative with respect to $%
\tilde{\beta}$ are continuous at half-integer values $\tilde{\beta}=\pm
1/2,\pm 3/2,\ldots $. This feature is seen from the right panel of Fig. \ref%
{fig4}. Concerning the FC as a function of the parameter $\alpha $
(containing the dependence on the magnetic flux through the cosmic string
core), it is continuous at $\alpha =\pm 1/2,\pm 3/2,\ldots $, but its
derivative with respect to $\alpha $ has discontinuities at those points.

Now we turn to the investigation of the FC in the asymptotic regions of the
values for the compactification length. For $L/w\gg 1$ and $L/r\gg 1$, we
can use again the asymptotic expression for the function ${\mathcal{Z}}%
_{ma}(u)$ given in (\ref{asymp1}). For the further estimate of the FC (\ref%
{FC-c}) two cases should be considered separately. For $2|\alpha _{0}|<1-1/q$%
, to the leading order, we can omit the parts $1+2\rho ^{2}\sin ^{2}(\pi
k/q) $ and $1+2\rho ^{2}\cosh ^{2}u$ in (\ref{args-c}) and this gives%
\begin{equation}
\langle \bar{\psi}\psi \rangle _{\mathrm{c}}\approx -\frac{2(2ma+1)(2ma+3)}{%
\pi ^{2}a^{4}(L/w)^{2ma+5}}\left[ 1+2h_{0}(q,\alpha _{0})\right]
\sum_{l=1}^{\infty }\frac{\cos (2\pi l\tilde{\beta})}{l^{2ma+5}}.
\label{FCclargeL}
\end{equation}%
This leading term does not depend on the radial coordinate $r$. In the range
of the parameters $2|\alpha _{0}|>1-1/q$ we cannot ignore $2\rho ^{2}\cosh
^{2}u$ with respect to $(lL/w)^{2}/2$ in the integral term (the integral
would diverge). This means that the dominant contribution to the integral in
(\ref{FC-c}) comes from large values of $u$ with $e^{u}\sim L/r$. By using
this fact it can be shown that for $2|\alpha _{0}|>1-1/q$ the contribution
of the integral term in (\ref{FC-c}) dominates and to the leading order the
condensate $\langle \bar{\psi}\psi \rangle _{\mathrm{c}}$ decays like $%
(w/r)^{2ma+5}(r/L)^{2ma+4+q-2|\alpha _{0}|q}$. In this case, the decay of
the compactification contribution as a function of $1/L$ is weaker. Note
that for the contribution $\langle \bar{\psi}\psi \rangle _{\mathrm{c}%
}^{(0)} $, by using (\ref{asymp1}) in (\ref{FCk0}), in the limit $L/w\gg 1$
one gets%
\begin{equation}
\langle \bar{\psi}\psi \rangle _{\mathrm{c}}^{(0)}\approx -\frac{%
2(2ma+1)(2ma+3)}{\pi ^{2}a^{4}(L/w)^{2ma+5}}\sum_{l=1}^{\infty }\frac{\cos
(2\pi l\tilde{\beta})}{l^{2ma+5}}.  \label{FCk0large}
\end{equation}%
This leading term coincides with the part in (\ref{FCclargeL}) coming from
the first term in the square brackets.

In order to find the asymptotic for small values of the ratio $L/w\ll 1$, it
is convenient to provide an alternative representation for the topological
part $\langle \bar{\psi}\psi \rangle _{\mathrm{t}}$. The latter is obtained
by combining the representations (\ref{FCcs0}) and (\ref{compc}) in (\ref%
{FCt}) and by using the relation 
\begin{equation}
\sum_{l=-\infty }^{\infty }e^{-l^{2}L^{2}x/2w^{2}+2\pi il\tilde{\beta}}=%
\frac{w\sqrt{2\pi }}{L\sqrt{x}}\sum_{l=-\infty }^{\infty }\exp \left[ -\frac{%
2w^{2}\pi ^{2}}{L^{2}x}\left( \tilde{\beta}-l\right) ^{2}\right] .
\label{Resum}
\end{equation}%
This relation directly follows from the Poisson resummation formula \cite%
{Pinsky2008}. The required representation reads%
\begin{eqnarray}
\langle \bar{\psi}\psi \rangle _{\mathrm{t}} &=&\langle \bar{\psi}\psi
\rangle _{\mathrm{c}}^{(0)}-\frac{\sqrt{2}w}{\pi ^{3/2}a^{4}L}%
\int_{0}^{\infty }dx\,x^{3/2}e^{-x}\left[ I_{\nu _{1}}(x)-I_{\nu _{2}}(x)%
\right] \sum_{l=-\infty }^{\infty }\exp \left[ -\frac{2w^{2}\pi ^{2}}{L^{2}x}%
\left( \tilde{\beta}-l\right) ^{2}\right]  \notag \\
&&\times \left[ \sideset{}{'}{\sum}_{k=1}^{[q/2]}(-1)^{k}\cos (\pi k/q)\cos
(2\pi k\alpha _{0})e^{-2x\rho ^{2}\sin ^{2}(\pi k/q)}\right.  \notag \\
&&\left. +\frac{q}{\pi }\int_{0}^{\infty }du\frac{h(q,\alpha _{0},2u)\sinh u%
}{\cosh (2qu)-\cos (q\pi )}e^{-2x\rho ^{2}\cosh ^{2}u}\right] \ .
\label{FCt2}
\end{eqnarray}%
Note that in the case of a massless field one has $I_{\nu _{1}}(x)-I_{\nu
_{2}}(x)=e^{-x}\sqrt{2/\pi x}$ and the integrals over $x$ in (\ref{FCt2})
are expressed in terms of the function $K_{2}(4\pi |\tilde{\beta}-l|\sqrt{%
w^{2}+r^{2}b^{2}}/L)$ with $b=\sin (\pi k/q)$ and $b=\cosh u$ for the parts
coming from the sum over $k$ and from the integral over $u$, respectively.
For the first term in the right-hand side of (\ref{FCt2}) we will use the
representation (the $k=0$ term in (\ref{compc})) 
\begin{equation}
\langle \bar{\psi}\psi \rangle _{\mathrm{c}}^{(0)}=-\frac{1}{\pi ^{2}a^{4}}%
\sum_{l=1}^{\infty }\cos (2\pi l\tilde{\beta})\int_{0}^{\infty
}dx\,x^{2}e^{-x}\left[ I_{\nu _{1}}(x)-I_{\nu _{2}}(x)\right]
e^{-x(lL)^{2}/2w^{2}}.  \label{FCk02}
\end{equation}%
First let us estimate this term for $L/w\ll 1$. Under this condition the
dominant contribution to (\ref{FCk02}) comes from large values of $x$. From
the corresponding asymptotic formulas for the modified Bessel function one
gets $I_{\nu _{1}}\left( x\right) -I_{\nu _{2}}\left( x\right) \approx
amx^{-3/2}e^{x}/\sqrt{2\pi }$ and to the leading order we obtain%
\begin{equation}
\langle \bar{\psi}\psi \rangle _{\mathrm{c}}^{(0)}\approx -\frac{2m(w/L)^{3}%
}{\pi ^{2}a^{3}}\sum_{l=1}^{\infty }\frac{\cos (2\pi l\tilde{\beta})}{l^{3}}.
\label{FCk0as}
\end{equation}%
The sum of the series in (\ref{FCk0as}) becomes zero for $\tilde{\beta}=%
\tilde{\beta}_{0}\approx 0.231$. The leading term (\ref{FCk0as}) is negative
for $|\tilde{\beta}|<\tilde{\beta}_{0}$ and positive for $\tilde{\beta}_{0}<|%
\tilde{\beta}|\leq 0.5$. The estimate of the last term in (\ref{FCt2})
essentially depends on $\tilde{\beta}$. The FC is a periodic function of $%
\tilde{\beta}$ with the period 1 and we consider the range $|\tilde{\beta}%
|<1/2$. For $\tilde{\beta}=0$ the main contribution gives the term $l=0$ and
the second term in the right-hand side behaves as $w/L$. In this case the
first term is estimated as $\langle \bar{\psi}\psi \rangle _{\mathrm{c}%
}^{(0)}\approx -2m(w/L)^{3}\zeta (3)/(\pi ^{2}a^{3})$ and it dominates in
the total FC. For $0<|\tilde{\beta}|<1/2$, again, the contribution of the
term $l=0$ dominates. Additionally assuming that $L\ll r$, we can see that
the main contribution to the integral over $x$ comes from the integration
range near $x\sim w/L$. Using the large argument asymptotic for the
difference of the modified Bessel functions, we can see that the last term
in (\ref{FCt2}) is suppressed by the factor $\left( Lr/w^{2}\right)
^{-3/2}\exp \left[ -4\pi |\tilde{\beta}|r\sin (\pi /q)/L\right] $ for $q\geq
2$ and by $\left( Lr/w^{2}\right) ^{-3/2}e^{-4\pi |\tilde{\beta}|r/L}$ for $%
1\leq q<2$. Hence, in all cases for small values of the compactification
length the FC (\ref{FCt2}) is dominated by the first term in the right-hand
side and it behaves like $(w/L)^{3}$.

\begin{figure}[tbph]
\begin{center}
\begin{tabular}{cc}
\epsfig{figure=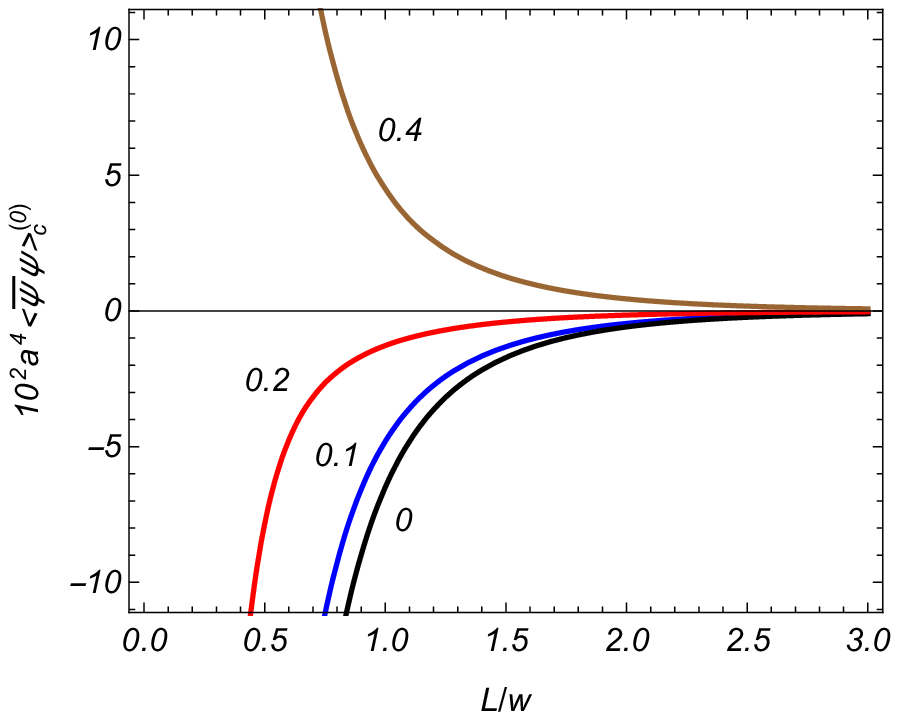,width=7.cm,height=5.5cm} & \quad %
\epsfig{figure=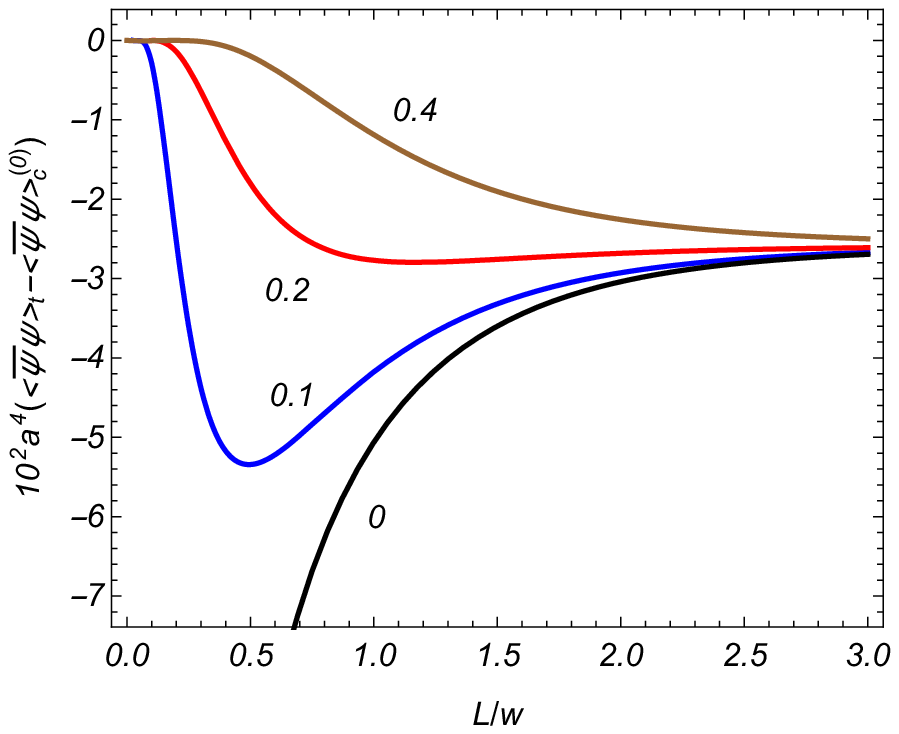,width=7.cm,height=5.5cm}%
\end{tabular}%
\end{center}
\caption{The contributions $\langle \bar{\protect\psi}\protect\psi \rangle _{%
\mathrm{c}}^{(0)}$ (left panel) and $\langle \bar{\protect\psi}\protect\psi %
\rangle _{\mathrm{t}}-\langle \bar{\protect\psi}\protect\psi \rangle _{%
\mathrm{c}}^{(0)}$ (right panel) in the FC as functions of the length of the
compact dimension. The numbers near the curves are the values of the
parameter $\tilde{\protect\beta}$. The values of the remaining parameters
are given in the text.}
\label{fig5}
\end{figure}

In Fig. \ref{fig5} we display the dependence of the FC on the proper length
of the compact dimension measured in units of the curvature radius. The
graphs are plotted for $q=2.5$, $ma=0.5$, $\alpha _{0}=0.3$, $r/w=1$, and
the numbers near the curves correspond to the values of the parameter $%
\tilde{\beta}$. The left panel presents the condensate $\langle \bar{\psi}%
\psi \rangle _{\mathrm{c}}^{(0)}$ for the special case of the problem with $%
q=1$ and $\alpha _{0}=0$ (AdS spacetime with compact dimension in the
absence of the cosmic string). On the right panel we have plotted the
difference $\langle \bar{\psi}\psi \rangle _{\mathrm{t}}-\langle \bar{\psi}%
\psi \rangle _{\mathrm{c}}^{(0)}$ (see (\ref{FCt1})) that presents the
effects induced by the cosmic string in AdS spacetime with a compact
dimension. For $\tilde{\beta}\neq \tilde{\beta}_{0}$, the condensate $%
\langle \bar{\psi}\psi \rangle _{\mathrm{c}}^{(0)}$ behaves like $(w/L)^{3}$
for small values of $L/w$. For large values of that ratio the FC\ decays as $%
(w/L)^{2ma+5}$ (see (\ref{FCk0large})). For small values of $L/w$ and for $%
\tilde{\beta}=0$ we get $\langle \bar{\psi}\psi \rangle _{\mathrm{t}%
}-\langle \bar{\psi}\psi \rangle _{\mathrm{c}}^{(0)}\propto w/L$, whereas
for $\tilde{\beta}\neq 0$ one has an exponential suppression. For large
values of $L/w$ the dominant contribution in $\langle \bar{\psi}\psi \rangle
_{\mathrm{t}}-\langle \bar{\psi}\psi \rangle _{\mathrm{c}}^{(0)}$ comes from
the term $l=0$ in (\ref{FCt1}) which coincides with the condensate $\langle 
\bar{\psi}\psi \rangle _{\mathrm{cs}}$, given by (\ref{FC-cs}). All these
features are confirmed by the numerical results in Fig. \ref{fig5}. 

Now we turn to the asymptotics near the AdS boundary and horizon. For points
near the AdS boundary one has $w\ll L,r$ and we use the representation (\ref%
{compc}). The dominant contribution to the integral over $x$ comes from the
region near the lower limit. By making use of the expansion for the modified
Bessel function for small values of the argument, for the leading term we get%
\begin{eqnarray}
\langle \bar{\psi}\psi \rangle _{\mathrm{c}} &\approx &-4\frac{\left(
2ma+3\right) \left( 2ma+1\right) }{\pi ^{2}a^{4}}w^{2ma+5}\sum_{l=1}^{\infty
}\cos (2\pi l\tilde{\beta})  \notag \\
&&\times \left[ \sideset{}{'_*}{\sum}_{k=0}^{[q/2]}\frac{(-1)^{k}\cos (\pi
k/q)\cos (2\pi k\alpha _{0})}{[l^{2}L^{2}+4r^{2}\sin ^{2}(\pi k/q)]^{ma+5/2}}%
\right.  \notag \\
&&\left. +\frac{q}{\pi }\int_{0}^{\infty }du\frac{h(q,\alpha _{0},2u)\sinh u%
}{\cosh (2qu)-\cos (q\pi )}\left( l^{2}L^{2}+4r^{2}\cosh ^{2}u\right)
^{-ma-5/2}\right] \ .  \label{FCcAdSb}
\end{eqnarray}%
This shows that the compactification part in the FC vanishes on the AdS
boundary as $w^{2ma+5}$. As it has been discussed in the previous section, a
similar behavior takes place for the contribution $\langle \bar{\psi}\psi
\rangle _{\mathrm{cs}}$ (see (\ref{larger})).

In the near-horizon region one has $w\gg L,r$, and we consider the cases of
massive and massless fields separately. Again, for a massive field we employ
the representation (\ref{compc}). Now the main contribution to the integral
over $x$ comes form the region with large values of $x$. For those $x$ one
has 
\begin{equation*}
I_{\nu _{1}}\left( x\right) -I_{\nu _{2}}\left( x\right) \approx \frac{%
mae^{x}}{\sqrt{2\pi }x^{3/2}},
\end{equation*}%
and the integrals over $x$ are expressed in terms of the gamma function. For
the leading order term we get 
\begin{eqnarray}
\langle \bar{\psi}\psi \rangle _{\mathrm{c}} &\approx &-\frac{2mw^{3}}{\pi
^{2}a^{3}}\sum_{l=1}^{\infty }\cos (2\pi l\tilde{\beta})\left[ %
\sideset{}{'_*}{\sum}_{k=0}^{[q/2]}\frac{(-1)^{k}\cos (\pi k/q)\cos (2\pi
k\alpha _{0})}{\left[ l^{2}L^{2}+4r^{2}\sin ^{2}(\pi k/q)\right] ^{3/2}}%
\right.  \notag \\
&&\left. +\frac{q}{\pi }\int_{0}^{\infty }du\frac{h(q,\alpha _{0},2u)\sinh u%
}{\cosh (2qu)-\cos (q\pi )}\left( l^{2}L^{2}+4r^{2}\cosh ^{2}u\right) ^{-3/2}%
\right] \ ,  \label{FCcAdSh}
\end{eqnarray}%
and near the horizon the compactification induced part behaves as $w^{3}$.

For a massless field we will use the formula (\ref{FCcm0}) for the
investigation of the near-horizon asymptotic. The condensate $\langle \bar{%
\psi}\psi \rangle _{\mathrm{c}}$ is a periodic function of the parameter $%
\tilde{\beta}$ with the period 1 and in that discussion we will assume $|%
\tilde{\beta}|\leq 1/2$. The asymptotic is different for $\tilde{\beta}\neq
0 $ and $\tilde{\beta}=0$ and we start with the first case. Under the
condition (\ref{cond1}) and near the horizon we can directly put $\rho =0$
in (\ref{FCcm0}). Then, the series over $l$ is estimated by using (\ref%
{seras}) with $b=1$. To the leading order this gives%
\begin{equation}
\langle \bar{\psi}\psi \rangle _{\mathrm{c}}\approx 3\frac{1+2h_{0}(q,\alpha
_{0})}{32\pi ^{2}a^{4}}.  \label{FCcAdShm0}
\end{equation}%
By taking into account that, under same conditions, one has $\langle \bar{%
\psi}\psi \rangle _{\mathrm{cs}}\approx -3h_{0}(q,\alpha _{0})/(4\pi
a^{2})^{2}$, for the topological part in the FC we get the asymptotic 
\begin{equation}
\langle \bar{\psi}\psi \rangle _{\mathrm{t}}\approx \frac{3}{32\pi ^{2}a^{4}}%
.  \label{FCtAdShm0}
\end{equation}%
Recall that it is obtained under the conditions $w\gg L,r$, $\tilde{\beta}%
\neq 0$, $m=0$ and in the range (\ref{cond1}) for the parameters. For $w\gg
L,r$, $\tilde{\beta}\neq 0$, $m=0$ and in the range (\ref{cond2}), the
contribution from the term containing the integral over $u$ dominates in (%
\ref{FCcm0}). The main contribution to the integral comes from large values
of $u$ and it can be seen that, in the leading order, $\langle \bar{\psi}%
\psi \rangle _{\mathrm{c}}\approx -\langle \bar{\psi}\psi \rangle _{\mathrm{%
cs}}$, where the asymptotic for $\langle \bar{\psi}\psi \rangle _{\mathrm{cs}%
}$ is given by (\ref{FCcsNear}). Hence, the leading contributions coming
from the separate terms in the right-hand side of the formula (\ref{FCt})
for the topological FC cancel each other. Related to that, for the
investigation of the behavior of the condensate $\langle \bar{\psi}\psi
\rangle _{\mathrm{t}}$ in the special case under consideration it is
convenient to use the representation (\ref{FCt2}). The dominant contribution
comes from the term $\langle \bar{\psi}\psi \rangle _{\mathrm{c}}^{(0)}$ in
the right-hand side and the result (\ref{FCtAdShm0}) is obtained for the
leading term. Hence, unlike the separate contributions $\langle \bar{\psi}%
\psi \rangle _{\mathrm{cs}}$ and $\langle \bar{\psi}\psi \rangle _{\mathrm{c}%
}$, for $\tilde{\beta}\neq 0$ the condensate $\langle \bar{\psi}\psi \rangle
_{\mathrm{t}}$ is finite in both regions (\ref{cond1}) and (\ref{cond2}).

It remains to consider the near-horizon behavior for a massless field with $%
\tilde{\beta}=0$. Under the condition (\ref{cond1}), we substitute $\rho =0$
in (\ref{FCcm0}) and then, by taking into account that the dominant
contribution to the series comes from large $l$, replace the summation by
the integration. In this way we can see that $\langle \bar{\psi}\psi \rangle
_{\mathrm{c}}\approx -w/h_{0}(q,\alpha _{0})/(2\pi ^{2}a^{4}L)$. Combining
this with the corresponding expression for $\langle \bar{\psi}\psi \rangle _{%
\mathrm{cs}}$ from the previous section, one finds%
\begin{equation}
\langle \bar{\psi}\psi \rangle _{\mathrm{t}}\approx -\frac{1+2h_{0}(q,\alpha
_{0})}{4\pi ^{2}a^{4}L}w.  \label{FCtAdShm0b}
\end{equation}%
In the range (\ref{cond2}) and for $\tilde{\beta}=0$ the leading
contribution in (\ref{FCcm0}) comes from the term involving the integral
over $u$. Estimating the latter in the way described above, we get%
\begin{equation}
\langle \bar{\psi}\psi \rangle _{\mathrm{c}}\approx \frac{q\left[ \left(
2|\alpha _{0}|-1\right) q-1\right] w}{16\pi ^{2}a^{4}L\left( \rho /2\right)
^{\left( 2|\alpha _{0}|-1\right) q+1}}.  \label{FCcAdShm0b}
\end{equation}%
In the same limit, the contribution $\langle \bar{\psi}\psi \rangle _{%
\mathrm{cs}}$ behaves as $1/\rho ^{\left( 2|\alpha _{0}|-1\right) q+1}$.
Hence, for a massless field with $\tilde{\beta}=0$ near-horizon asymptotic
is dominated by the contribution induced by the compactification and $%
\langle \bar{\psi}\psi \rangle _{\mathrm{t}}\approx \langle \bar{\psi}\psi
\rangle _{\mathrm{c}}$ with $\langle \bar{\psi}\psi \rangle _{\mathrm{c}}$
given in (\ref{FCcAdShm0b}). In this case the topological part in the FC
behaves like $w^{2-\left( 1-2|\alpha _{0}|\right) q}$.

Let us recall that the investigation of the effects induced by the
compactification of the $z$-direction was presented in this Section in terms
of the fields $\psi _{(s)}^{\prime }$ with the Lagrangian density (\ref{Lsp}%
). This means that the topological part we have discussed above corresponds
to the condensate $\langle \bar{\psi}_{(s)}^{\prime }\psi _{(s)}^{\prime
}\rangle _{\mathrm{c}}$. As it has been shown, it has opposite signs for the
fields with $s=+1$ and $s=-1$. Returning to the initial representation with
the fields $\psi _{(s)}$, having the Lagrangian density (\ref{Ls}), and by
taking into account the relation $\langle \bar{\psi}_{(s)}\psi _{(s)}\rangle
_{\mathrm{c}}=s\langle \bar{\psi}_{(s)}^{\prime }\psi _{(s)}^{\prime
}\rangle _{\mathrm{c}}$, we conclude that the compactification contributions 
$\langle \bar{\psi}_{(s)}\psi _{(s)}\rangle _{\mathrm{c}}$ coincide for the
fields realizing two inequivalent irreducible representations of the
Clifford algebra. In particular, the FC in the models invariant under the
parity transformation and charge conjugation is obtained from the results
given in this section with an additional coefficient 2.

In the model we have considered the only interactions of the fermionic field
are with the background gravitational and electromagnetic fields. The FC is
an important quantity in theories involving fermions interacting with other
quantum fields. In particular, the FC appears as an order parameter that
governs the phase transitions in those theories. The results obtained in the
present paper can be considered as the first step in considering the
combined topological effects of cosmic strings and compactification in
interacting theories. The topological contributions in the FC may lead to
interesting effects such as the topological mass generation, symmetry
restoration and instabilities. For example, in models involving a scalar
field $\varphi (x)$, with the interaction term in the Lagrangian density
proportional to $\varphi ^{2}\bar{\psi}\psi $, the formation of the nonzero
FC leads to the term in the equation for the scalar field that is
proportional to $\varphi \langle \bar{\psi}\psi \rangle $. This leads to the
shift in the effective mass for $\varphi (x)$ determined by the FC (for a
similar discussion for two interacting scalar fields see, e.g., \cite%
{Ford80,Toms80}). To the leading order with respect to the scalar-fermion
interaction the mass shift is determined by the FC evaluated within the
framework of the free-fermion model. Similar features may appear in
Nambu-Jona-Lasinio-type four-fermion models with the self-interaction $(\bar{%
\psi}\psi )^{2}$ (see, for example, \cite{Klim88}-\cite{Flac13}). Again, to
the leading order, the shift in the fermionic mass is determined by the
condensate we have discussed above. Depending on the sign, the topological
shift in the FC may lead to the restoration of the symmetries or to
instabilities in interacting field theories.

\section{Conclusion}

\label{sec:Conc}

In the present paper we have investigated combined effects of the
gravitational field and spatial topology on the FC in (1+4)-dimensional
spacetime. In order to have an exactly solvable problem, a highly symmetric
background spacetime is considered with the locally AdS geometry. The
nontrivial toplogy is implemented by the compactification of the $z$
coordinate and the presence of a cosmic string carrying a magnetic flux. For
points outside the string's core the influences of the cosmic string and
compactification are purely topological. In odd dimensional spacetimes one
has two inequivalent irreducible representations of the Clifford algebra. In
order to unify the investigation of the FCs for the corresponding fields, we
have passed to a new representation where the Dirac equations for those
fields differ by the sign in front of the mass term.

First we have discussed the geometry where the $z$-direction has trivial
topology. The contribution induced in the FC by the cosmic string is
expressed as (\ref{FC-cs}), where the function ${\mathcal{Z}}_{ma}(u)$ is
given in terms of the associated Legendre function of the second kind. This
contribution is an even periodic function of the magnetic flux inside the
string core with the period equal to the flux quantum. The general
expression for the string induced part, $\langle \bar{\psi}\psi \rangle _{%
\mathrm{cs}}$, is further simplified to (\ref{FCm0}) for a massless field.
By the limiting transition we have obtained the FC around a cosmic string in
(1+4)-dimensional Minkowski spacetime. For a massive field the latter is
given by (\ref{FCcsM}) and it vanishes for a massless field. This shows that
the nonzero FC on the AdS bulk for massless fermionic fields is the effect
induced by the gravitational field. For massive fields and at small proper
distances from the string, $r_{p}\ll a$, the leading term in the asymptotic
expansion for $\langle \bar{\psi}\psi \rangle _{\mathrm{cs}}$ is given by (%
\ref{Fccsrsm}) and it diverges on the string as $1/r_{p}^{3}$. For a
massless field on the AdS bulk the condensate $\langle \bar{\psi}\psi
\rangle _{\mathrm{cs}}$ is finite on the string under the condition (\ref%
{cond1}) and diverges like $1/\rho ^{1+\left( 2|\alpha _{0}|-1\right) q}$ in
the range of parameters $2|\alpha _{0}|>1-1/q$. At large distances from the
string the contribution $\langle \bar{\psi}\psi \rangle _{\mathrm{cs}}$
decays as $(w/r)^{2ma+5}$. Unlike the case of the Minkowski bulk, the
fall-off in AdS bulk is power-law for both massless and massive fields.

The effects induced by the compactification of the $z$-direction are studied
in Section \ref{sec:Comp}. The corresponding contribution in the FC is
explicitly separated by using the summation formula (\ref{AP}). For general
case of a massive field it is given by the expression (\ref{FC-c}). For a
massless field the corresponding expression is further simplified to (\ref%
{FCcm0}). Because of the maximal symmetry of the AdS spacetime, the
dependence of the compactification part on the variables $r$, $L$, $w$
enters in the form of the ratios $r/w$ and $L/w$. The latter is the proper
length of the compact dimensions in units of the curvature radius, measured
by an observer with a given value of the coordinate $w$. In the absence of
the planar angle deficit and of the magnetic flux the term $k=0$, given by (%
\ref{FCk0}), survives only. The remaining part contains the effects induced
by the cosmic string and magnetic flux. For both massive and massless field
the compactification contribution in the FC is finite on the string for $%
2|\alpha _{0}|<1-1/q$ and diverges as $1/\rho ^{1+\left( 2|\alpha
_{0}|-1\right) q}$ under the condition (\ref{cond2}). This behavior is
similar to that for the part $\langle \bar{\psi}\psi \rangle _{\mathrm{cs}}$
in the case of a massless field. As a limiting case, we have obtained the
compactification contribution for a cosmic string in the Minkowski bulk. For
a massive field it is given by the formula (\ref{FCcM}) and vanishes for a
massless field. Hence, for a cosmic string in background of the
(1+4)-dimensional Minkowski spactime with compactified $z$-direction the
total FC vanishes for a massless field.

For the AdS bulk and at large proper distances from the string the FC tends
to the part $\langle \bar{\psi}\psi \rangle _{\mathrm{c}}^{(0)}$ and the
contribution induced by the planar angle deficit and magnetic flux decays as 
$(w/r)^{2ma+5}$ (see (\ref{FCclarger})). It is of interest to note that the
leading terms in the expansions of the parts $\langle \bar{\psi}\psi \rangle
_{\mathrm{cs}}$ and $\langle \bar{\psi}\psi \rangle _{\mathrm{c}}$ cancel
each other and the decay of the contribution from planar angle deficit and
magnetic flux in the total FC is stronger. For large values of the length of
the compact dimension and under the condition (\ref{cond1}) the
compactification contribution decays as $(L/w)^{2ma+5}$ and the leading term
in the corresponding asymptotic expansion does not depend on the radial
coordinate. In the range $2|\alpha _{0}|>1-1/q$ and for large values of the
compactification length the contribution $\langle \bar{\psi}\psi \rangle _{%
\mathrm{c}}$ behaves as $(w/r)^{2ma+5}(r/L)^{2ma+4+q-2|\alpha _{0}|q}$. In
order to investigate the FC for small values of the compactification radius
we have provided an alternative representation (\ref{FCt2}) for the
topological contribution. In this limit the FC $\langle \bar{\psi}\psi
\rangle _{\mathrm{t}}$ is dominated by the part $\langle \bar{\psi}\psi
\rangle _{\mathrm{c}}^{(0)}$ and it behaves like $(w/L)^{3}$. The part $%
\langle \bar{\psi}\psi \rangle _{\mathrm{t}}-\langle \bar{\psi}\psi \rangle
_{\mathrm{c}}^{(0)}$ is induced by the cosmic string and by the magnetic
flux. Its behavior for small compactification radius crucially depends
whether the parameter $\tilde{\beta}$, $|\tilde{\beta}|<1/2$, is zero or
not. For $\tilde{\beta}=0$ one has $\langle \bar{\psi}\psi \rangle _{\mathrm{%
t}}-\langle \bar{\psi}\psi \rangle _{\mathrm{c}}^{(0)}\propto w/L$ and the
last term in the right-hand side of (\ref{FCt2}) is large, though
subdominant to compared with the contribution coming from $\langle \bar{\psi}%
\psi \rangle _{\mathrm{c}}^{(0)}$. For $0<|\tilde{\beta}|<1/2$ the effects
induced by the cosmic string and by the magnetic flux corresponding to the
parameter $\alpha _{0}$ are suppressed exponentially, by the factor $\exp
[-4\pi |\tilde{\beta}|r\sin (\pi /q)/L]$ for $q\geq 2$ and by $e^{-4\pi |%
\tilde{\beta}|r/L}$ in the range $1\leq q<2$. We note that in the special
case of the absence of planar angle deficit, corresponding to $q=1$, the
results presented in this paper describe Aharonov-Bohm-type effects induced
by magnetic fluxes in the AdS spacetime. The separate contribution to the FC
for this special case are given by the formulas (\ref{FCmf}) and (\ref{FCcq1}%
).

Both the contributions in the FC, $\langle \bar{\psi}\psi \rangle _{\mathrm{%
cs}}$ and $\langle \bar{\psi}\psi \rangle _{\mathrm{c}}$, tend to zero on
the AdS boundary like $w^{2ma+5}$. Their behavior is more diversified near
the horizon, corresponding to large values of the coordinate $w$. For a
massive field the separate contributions behave as $w^{3}$. The leading
terms in the corresponding asymptotic expansions are proportional to the
mass. For a massless field these leading terms vanish and the near-horizon
behavior of the FC is essentially different for $0<|\tilde{\beta}|\leq 1/2$
and for $\tilde{\beta}=0$. In the first case, both the parts are finite on
the horizon and the limiting value for the topological condensate $\langle 
\bar{\psi}\psi \rangle _{\mathrm{t}}$ is expressed by (\ref{FCtAdShm0}). For 
$\tilde{\beta}=0$ and in the range of parameters (\ref{cond1}), the leading
term in the near-horizon expansion is given by (\ref{FCcAdShm0b}) and the
condensate behaves like $\langle \bar{\psi}\psi \rangle _{\mathrm{t}}\propto
w$. Under the condition (\ref{cond2}) the divergence on the horizon is
stronger, as $w^{2-\left( 1-2|\alpha _{0}|\right) q}$.

\section*{Acknowledgments}

A.A.S. was supported by the grant No. 20RF-059 of the Committee of Science
of the Ministry of Education, Science, Culture and Sport RA. E.R.B.M. is
partially supported by CNPQ under Grant no. 301.783/2019-3. W.O.S. thanks
Coordena\c{c}\~{a}o de Aperfei\c{c}oamento de Pessoa de N\'{\i}vel Superior
(CAPES) for financial support.

\end{document}